\documentclass[9pt,twocolumn,twoside]{osajnl}
\journal{jocn} 

\usepackage{natbib}
\usepackage[flushleft]{threeparttablex}
\usepackage{todonotes}
\usepackage{amsmath}
\usepackage{bm}
\setboolean{shortarticle}{false}

\usepackage{orcidlink}

\usepackage{hyperref}

\newcommand{\oid}[1]{\orcidlink{#1}\,}
\renewcommand{\S}{\mathcal{S}}
\newcommand{\N}{\mathcal{N}}
\newcommand{\C}{\mathcal{C}}
\newcommand{\rabs}{\bar{|r|}}
\newcommand{\Exp}[1]{\left\langle{#1}\right\rangle}

\AtBeginDocument{}

\title{Availability, outage, and capacity of spatially correlated, Australasian free-space optical networks}

\author[1,*]{Marcus Birch$^{\oid{0000-0002-6215-5900}}$}
\author[2,3,$\dagger$]{James R. Beattie$^{\oid{0000-0001-9199-7771}}$}
\author[1]{Francis Bennet}
\author[4]{Nicholas Rattenbury}
\author[1]{Michael Copeland}
\author[1]{Tony Travouillon}
\author[5]{Kate Ferguson}
\author[6]{John Cater}
\author[4]{Mikhael Sayat}

\affil[1]{Advanced Instrumentation Technology Centre, Research School of Astronomy and Astrophysics, Australian National University, Canberra, ACT 2611, Australia}
\affil[2]{Research School of Astronomy and Astrophysics, Australian National University, Canberra, ACT 2611, Australia}
\affil[3]{Department of Astronomy and Astrophysics, University of California, Santa Cruz, 1156 High Street, Santa Cruz, CA 96054, United States of America}
\affil[4]{Te Pūnaha Ātea - Space Institute, The University
of Auckland, Auckland 1010, New Zealand}
\affil[5]{ANU Institute for Space, Australian National University, Canberra, ACT 2601, Australia}
\affil[6]{Aerospace Research Centre, University of Canterbury, Christchurch 8140, New Zealand}

\affil[*]{Corresponding author: marcus.birch@anu.edu.au}
\affil[$\dagger$]{Corresponding author: james.beattie@anu.edu.au}

\begin{abstract}
 Network capacity and reliability for free space optical communication (FSOC) is strongly driven by ground station availability, dominated by local cloud cover causing an outage, and how availability relations between stations produce network diversity. We combine remote sensing data and novel methods to provide a generalised framework for assessing and optimising optical ground station networks. This work is guided by an example network of eight Australian and New Zealand optical communication ground stations which would span approximately $60^\circ$ in longitude and $20^\circ$ in latitude. Utilising time-dependent cloud cover data from five satellites, we present a detailed analysis determining the availability and diversity of the network, finding the Australasian region is well-suited for an optical network with a 69\% average site availability and low spatial cloud cover correlations. Employing methods from computational neuroscience, we provide a Monte Carlo method for sampling the joint probability distribution of site availabilities for an arbitrarily sized and point-wise correlated network of ground stations. Furthermore, we develop a general heuristic for site selection under availability and correlation optimisations, and combine this with orbital propagation simulations to compare the data capacity between optimised networks and the example network. We show that the example network may be capable of providing tens of terabits per day to a LEO satellite, and up to 99.97\% reliability to GEO satellites. We therefore use the Australasian region to demonstrate novel, generalised tools for assessing and optimising FSOC ground station networks, and additionally, the suitability of the region for hosting such a network.
\end{abstract}

\setboolean{displaycopyright}{true}

\begin{document}
\maketitle

\section{Introduction}
\label{sec:introduction}

\begin{figure*}[!ht]
  \centering
  \includegraphics[width=\textwidth]{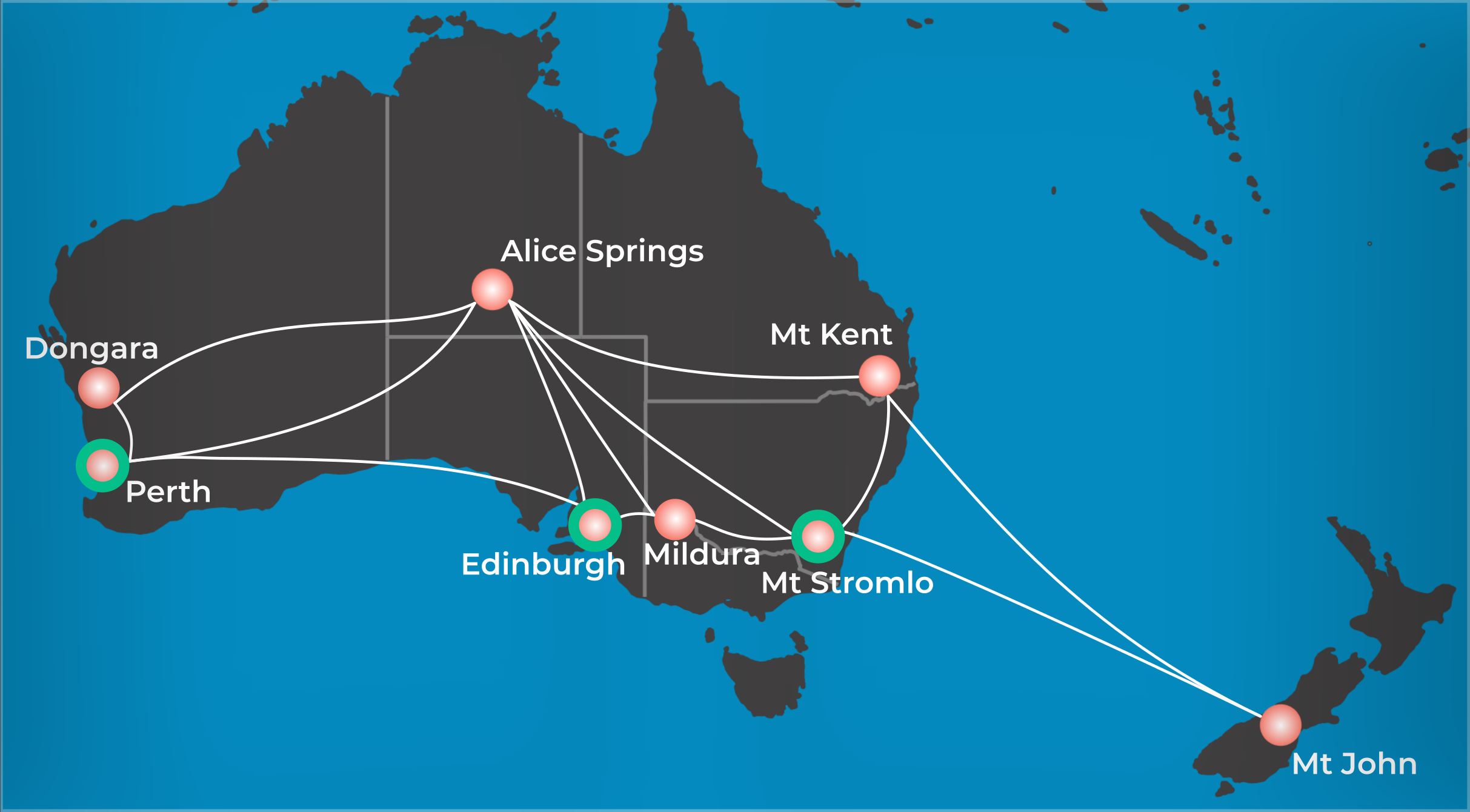}
  \caption{Australasian optical ground stations currently being commissioned (green circles; ``the base network"), and proposed for this study (red circles; ``the extended network"), comprising the Australasian Optical Ground Station Network (AOGSN). Throughout our analysis we use both of these networks as case studies for availability, correlation, and network optimisation in the Australasian region.}
  \label{fig:node_map}
\end{figure*}

Australia and New Zealand are pursuing a network of optical ground stations (OGS) to support the future of optical communication in the region~\cite{bennet2020australia}. Free space optical communication (FSOC) offers myriad advantages over existing radio frequency technologies, but outages due to clouds are a long-acknowledged barrier to its widespread adoption and reliability \cite{piazzolla2002statistics}. For satellite operators seeking to transfer large quantities of data in a timely manner, an extensive and reliable ground segment is critical. The typical solution to cloud-induced outages is operating a network of OGS and building redundancy into the system by providing site diversity such that it is improbable all OGS are obscured by cloud \cite{erdogan2021site,del2017optimal,net2016approximation,lyras2019medium}. In this paper we aim to develop a means of evaluating the capability of an FSOC network, including availability, diversity, and ultimately data capacity, alongside a general framework for network optimisation.

A well-optimised FSOC network is one with high availability at each site, and minimal correlation of cloud cover between sites for maximising diversity. Low Earth orbit (LEO) satellites can improve data throughput by engaging with a well-optimised network, increasing the number of possible links. For Geostationary orbit (GEO) satellites and deep space, where mutual visibility can be simultaneously maintained to multiple OGS, a well-optimised network boosts reliability, i.e. the probability that at least one OGS is available.

We approach this issue by using the proposed Australasian Optical Ground Station Network (AOGSN) as an example \cite{bennet2020australia}. The AOGSN will include at least the Australian National University's Optical Communication Ground Station at Mount Stromlo Observatory, the Australian Defence Science \& Technology Group OGS in Edinburgh, South Australia, and the University of Western Australia's OGS in Perth \cite{bennet2020australia,walsh2021western,birch2022mount,mudge2022dstg}. All three of these are in a state of commissioning and aim to enter operation in 2023. To create a robust model network to study, we propose an additional five sites at Mt Kent Observatory in Queensland, Mt John Observatory in New Zealand, Alice Springs in the Northern territory, Mildura in Victoria, and Dongara in central Western Australia. These additional sites are chosen as places with existing observatories or satellite ground segment infrastructure, and represent realistic OGS placements. The base and extended AOGSN will be used henceforth for reference to the initial three OGS as a network or all eight OGS respectively. \autoref{fig:node_map} shows the sites of this proposed network. We use the example of the extended AOGSN throughout this paper to compare with a network chosen from an optimisation model, and to test methods for availability, diversity, and capacity estimation.

Complex theoretical models have been developed for determining and predicting outages, including factors such as fading of the atmospheric channel due to turbulence \cite{erdogan2021site}. An experimental campaign to measure atmospheric turbulence at selected sites is considered in Section~\ref{sec:conclusion}, however this paper focuses on a spatial availability model without prior site selection. Therefore, as with earlier literature, we adopt a simple model for site outages: An OGS experiences an outage only due to cloud cover, which prevents the ground station from linking with the orbiting satellite \cite{fuchs2015ground,lyras2018optimum}.

Unlike clouds, where associated outages are a generic issue for any network, turbulence-induced outages are strongly system-specific. They may depend upon the orbital height of the satellite, details of the telescope and ground station specifications, e.g adaptive optics and uplink or downlink capabilities, etc \cite{martinez2018toward,osborn2021adaptive,schieler2023orbit,roberts2023performance}. This means that our generic, cloud-based outage model will require careful comparison with a real system, on a case-by-case basis. For example, an adaptive optics-compensated downlink from GEO, at a high quality site, is unlikely to experience many turbulent outages and will therefore have availability close to the cloud availability. Alternatively, uplinking to a LEO satellite in the same atmospheric conditions could suffer turbulent degradation that prevents the AO system from closing the loop for long periods of time, causing an outage. The difficulties of predicting turbulence conditions, and how this could be expanded into a rigorous turbulent and cloud availability model are discussed further in Section~\ref{sec:conclusion}. Our current model therefore encapsulates availability as the probability that a site is not cloudy at any given time. Given atmospheric turbulence might create additional outages, our model provides an upper bound on a realistic availability, which may be required for a detailed system-specific network performance analysis, which is beyond the scope of this study.

We provide a more rigorous definition of this cloud model based on the Bernoulli distribution in \autoref{sec:cover}. Remote sensing satellites measure cloud cover in different ways and with different issues, prompting us to adopt an approach that incorporates numerous satellites at different temporal and spatial resolutions, with decades of data in time. Deriving network diversity requires knowledge of the point-wise cloud correlation between sites. Simultaneous cloud cover measurements of the entire Australasian region are only possible with GEOs given their coverage, therefore this study places emphasis on data acquired from Himawari-8, a GEO weather satellite operated by the Japan Meteorological Agency, throughout our analysis \cite{bessho2016introduction}. 

\begin{figure*}[!ht]
  \centering
  \includegraphics[width=0.7\textwidth]{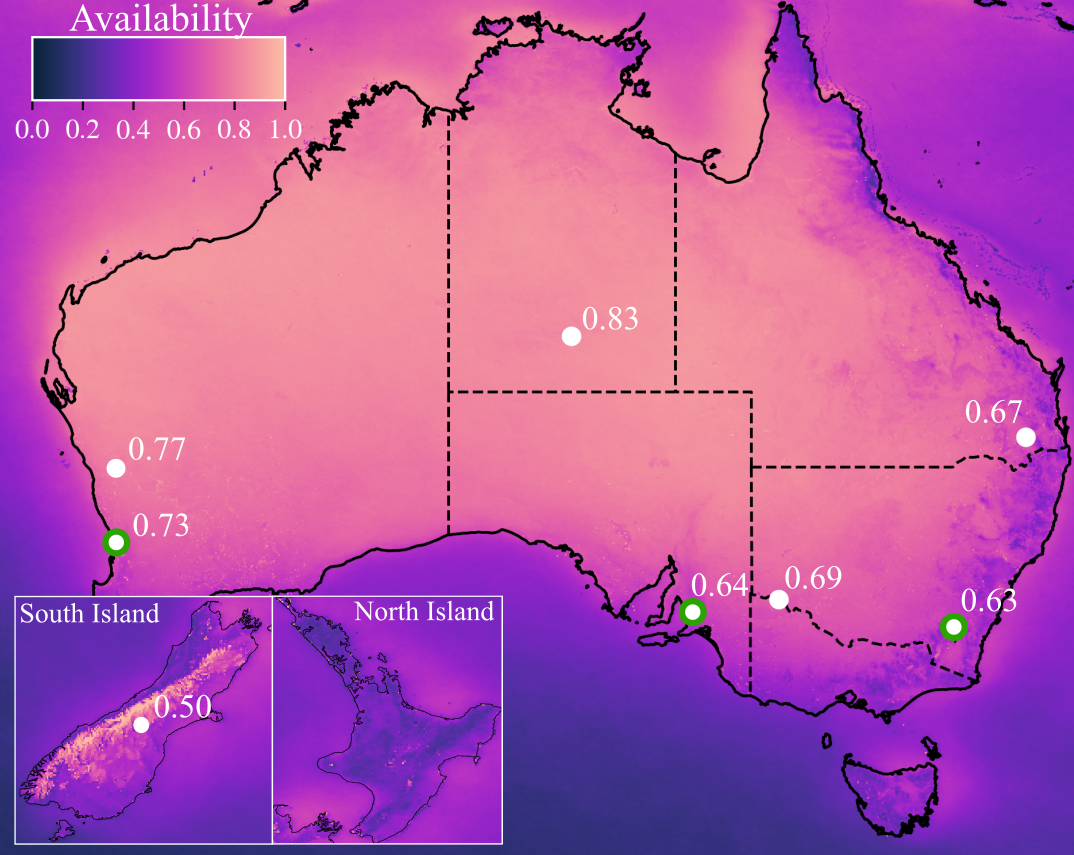}
  \caption{Availability of Australasian region as detected by AHI/Himawari-8, averaged from using twice daily cloud cover masks from 2015 to 2022. White dots correspond to geographical position of all OGS of the extended AOGSN, and green outlines indicate base AOGSN sites. Annotated availabilities are shown for each site. Data provided by Earth Observation Research Centre (Japan Aerospace Exploration Agency).}
  \label{fig:ANZ_clouds}
\end{figure*}

We develop a general method for computing outage probabilities for an arbitrary sized network with arbitrary point-wise correlations between each node, analogous to a network of spiking biological neurons with a specified covariance structure~\cite{Macke2009_neuronal_spike_model}. We sample our model using the availability and point-wise correlation from the base and extended AOGSN to provide robust outage statistics, extending the analysis performed in \cite{fuchs2015ground}. Furthermore, utilising our availability and diversity statistics we compare the base and extended AOGSN outage probabilities with a network that is optimised to maximise availability and minimise point-wise correlation for every ground station (node). Diversity optimisation methods have been recently developed~\cite{erdogan2021site,lyras2019medium,lyras2018optimum,poulenard2015ground}. However, we present a novel model for network optimisation that is spatially-resolved, not requiring pre-selected site locations.

This study is organised by primarily using the AOGSN model network as an example to drive methods and provide realistic outputs, and is structured as follows. In \autoref{sec:cover} we measure cloud cover over the entire Australasian region, and for sites of the AOGSN. \autoref{sec:correlation} covers the use of high temporal resolution, and simultaneous cloud measurements to directly measure spatial correlation of cloud cover, and correlations between selected sites. We then present a novel Monte Carlo method in \autoref{sec:div} for determining the outage probabilities of a network based on the availabilities and correlations between ground stations. A simple heuristic model for site selection to maximise availability and diversity is presented in \autoref{sec:optimisation}, then used to estimate an optimised network which is compared with the AOGSN across a number of metrics. Finally, in \autoref{sec:coverage}, we use orbital simulations to estimate the coverage of the network, and statistical link duration per day with LEO satellites. Estimates of network-wide link duration from simulation are combined with availabilities to estimate network capacity. For GEOs, we use our outage probability model to determine the network outage probability as a function of orbital longitude in \autoref{sec:geo_visibility}. Finally, in \autoref{sec:conclusion}, we summarise and list the key results of the study.

\section{Cloud cover and site availability}
\label{sec:cover}
Cloud cover, $\C$, is the major factor in degrading an FSOC network's availability, $A$ \cite{piazzolla2002statistics,venkat2013cloud}. Although satellites measure $\C$ in different ways, and many data products we use have already undergone spatial or time averaging, they are all capable of producing a binary cloud mask which flags $\C$ for each pixel, $(x_i,y_j)$, and time, $t_n$, in a satellite image. We consider that $\C$ is instantaneously measured as 0 (no cloud detection) or 1 (cloud detection) for each $(x_i,y_j)$ and $t_n$. Hence, over a time-interval, $t\in [t_1,t_n]$, the cloud cover is a vector $\hm{C}$ of $\left\{0,1\right\}$, with length $n$. Each component of $\hm{C}=(C_1,C_2,\hdots C_{n})$, $\C_n$ follows a Bernoulli distribution, $\C_n \sim  \text{Bernoulli}(\Omega)$, where $p(\C_n = 1) = \Omega$, and $p(\C_n = 0) = 1-\Omega = A$, hence,
\begin{align} \label{eq:Bernoulli_dis}
    p(\C;\,\Omega) = \Omega^{\C}A^{1-\C},
\end{align}
is the probability distribution function for $\C$. This is true for each $(x_i,y_j)$ on the grid. 

Let us consider a network $\N$ with $N$ nodes or sites. We call the $k^{\text{th}}$ site $\S_k$, and hence $\N\equiv \left\{ \S_1, \S_2, \hdots, \S_N\right\}$ is the set of $N$ sites, each with grid coordinates $(x_i,y_j)$. We may choose to index the sites themselves, rather than the grid coordinates, e.g., the probability distribution function at site $k$ is $p_k(\C_k;\Omega_k)$\footnote{We avoid a subscript of $\C$ for the time vector component index, $n$, which would interfere with our $k$ site index for simplicity.}. Because at any point in time each grid coordinate follows a Bernoulli distribution for $\C$, so does each site. Therefore, the probability density function for each $k$ site is $p_k(\C;\Omega_k) = \Omega_k^{\C_k}A_k^{1-\C_k}$. Note, directly from \autoref{eq:Bernoulli_dis} we can immediately estimate the availability $A$ at site $k$ through the time-average\footnote{We use the ensemble $\Exp{X}_t$ notation for the time-average of the quantity $X$.} $\Exp{\bm{\C}_k}_t=\Omega_k=1-A_k$.

$A$ has the most intuitive interpretation for FSOC ground stations as it corresponds to the probability a site is available for communications. We conduct an extensive study of $\C$ from numerous satellites to determine $A$ for each ground station of the AOGSN.  Five separate remote sensing satellites and instruments are used to determine $A$, and long time spans are used to minimise climatic variation. \autoref{tab:sats} summarises the satellites and instruments used for this study and the relevant time span of data associated with each. We use derived cloud retrieval products, with a focus on Sun-synchronous LEO satellites, and Himawari-8, which has been the prime GEO for use over the Australasian region since 2015~\cite{bessho2016introduction}.

Although Earth observation LEO satellites can produce cloud mask swathes with $<100$~m spatial resolution, we find working with low resolution global averages easier given the decades long time spans involved and subsequent quantity of data. Comparisons with higher resolution data products, e.g. from Himawari-8, found reasonable agreement, likely from the large time span of data involved and the relatively homogeneous terrain of Australia (\autoref{sec:satellitevariation}). New Zealand is naturally more impacted by spatial averaging; \autoref{sec:seasonalvariation} outlines how \textit{Aqua}/AIRS and MERRA-2 data is discarded for the Mt John node in the AOGSN. Therefore, for the LEO satellites: \textit{Terra}+\textit{Aqua}/MODIS, \textit{Aqua}/AIRS, and Suomi-NPP/VIIRS data, we use $1^\circ\times1^\circ$ global grids with monthly $\Omega$ instead of raw $\C$ measurements at high resolution~\cite{platnick2016modis,aumann2003airs,platnick2020nasa,frey2020continuity}.

Spatial averaging for terrain more varied than the Australian continent is likely to induce more errors, so doing the full computation on individual swathes would be preferred. MERRA-2, a modern retrospective analysis of assimilated satellite data since 1980, is able to provide a uniquely long time span with daily cloud information at $0.5^\circ\times0.6^\circ$ spatial resolution~\cite{gelaro2017modern}. However, MERRA-2's reliance on older satellite data and large scale spatial averaging and interpolation can produce erroneous results. MERRA-2 is analysed before including in any statistical means (e.g. for the Mt Stromlo and Alice Springs station it produces outlier results compared with other data). Long time span data, such as MERRA-2, is important for reducing uncertainty from periodic climatological phenomena such as the El Niño Southern Oscillation and Indian Ocean Dipole~\cite{england2006interannual}. However, anthropogenic climate change is a non-periodic climatogolical effect that can induce significant error on long time span data, e.g. MERRA-2, as the effect on clouds and their circulation has been studied extensively~\cite{norris2016evidence,tselioudis2016midlatitude,choi2014further,nguyen2015expansion}.

\begin{table}[!ht]
    \centering
        \caption{Remote sensing satellites used in this paper, and spatial resolution of relevant cloud retrieval data products used.}
        \label{tab:results_summary}
            \begin{tabular}{p{2cm} | p{2cm} | p{1cm} | p{2cm}}
                \toprule
                Satellite & Instrument & Time span & Spatial Resolution\\
                \hline\hline
                Himawari-8 & AHI & 2015- & 5~km\\
                \hline
                \textit{Terra}/\textit{Aqua} & MODIS & 2002- & $1^\circ$\\
                \hline
                \textit{Aqua} & AIRS & 2003-& $1^\circ$\\
                \hline
                Suomi NPP & VIIRS & 2012- & $1^\circ$\\
                \hline
                MERRA-2 & \textit{Assimilation} & 1980- & $0.5^\circ$\\
                \hline\hline
            \end{tabular}
    \begin{tablenotes}
        \item \textit{\textbf{Notes.}} \textit{MERRA-2 is a historical assimilation of data from numerous sources provided by the NASA Global Modelling and Assimilation Office~\cite{gelaro2017modern}.}
    \end{tablenotes}
    \label{tab:sats}
\end{table}

\subsection{Spatially resolved Himawari-8 maps}

Himawari-8 is used for establishing an initial, spatially-resolved picture of $A$ over the Australasian region, given it is a GEO satellite that produces a cloud mask of the entire region every ten minutes at a 5~km resolution~\cite{bessho2016introduction}. Due to the large quantity of data produced by Himawari-8 since 2015, images at 9am and 3pm Australian Eastern Standard Time (UTC+10) are used, an Australian convention established by the Bureau of Meteorology. \autoref{fig:ANZ_clouds} shows $A$ over eight years of Himawari-8 daytime cloud product provided by JAXA/EORC\footnote{Japan Aerospace Exploration Agency, Earth Observation Research Centre} and the Japan Meteorological Agency~\cite{ishida2009development,iwabuchi2018cloud}. Cirrus cloud identified by Himawari-8 is considered to be clear sky for \autoref{fig:ANZ_clouds} given its low $1.55$~\textmu m attenuation~\cite{awan2009cloud}. Himawari-8 can produce misleading results for pixels with high altitude areas that typically sit above cloud height, seen for the Alps down the length of New Zealand's South Island in \autoref{fig:ANZ_clouds}~\cite{iwabuchi2018cloud}, but this does not affect any area of Australia.

\subsection{Seasonal variation with collated data}
\label{sec:seasonalvariation}

\begin{figure}
  \centering
  \includegraphics[width=\linewidth]{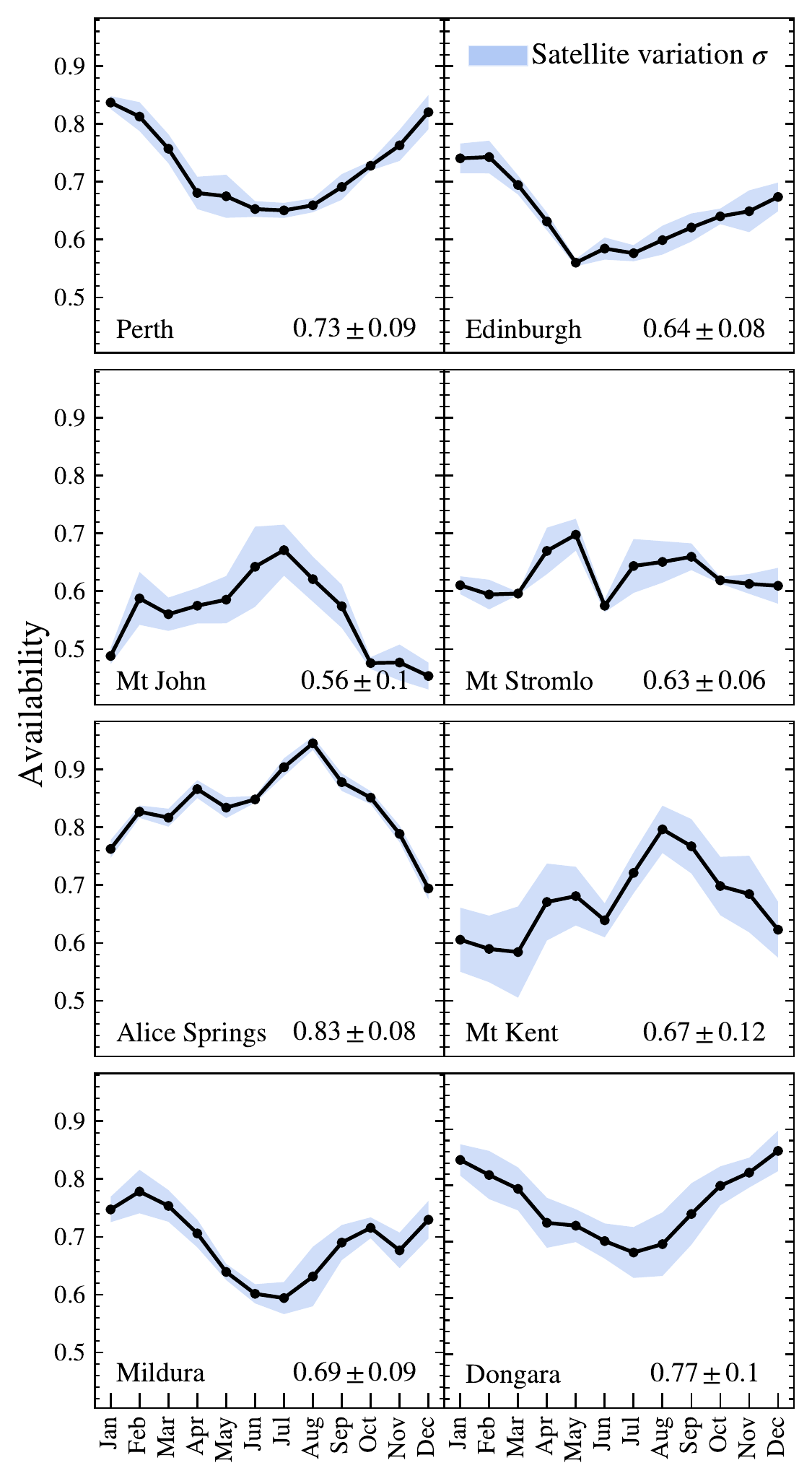}
  \caption{Seasonal availability for all ground stations in extended AOGSN. The availability, $A$, is averaged from satellites tabulated in \autoref{tab:sats}. The $1\sigma$ fluctuations, visualised with the blue band, display the variability between the different satellites. Uncertainty in the annual average of $A$ combines the $\sigma$ from satellite and the $\sigma$ over the monthly values.}
  \label{fig:aogsn_seasonal}
\end{figure}

\begin{figure*}[!ht]
  \centering
  \includegraphics[width=\textwidth]{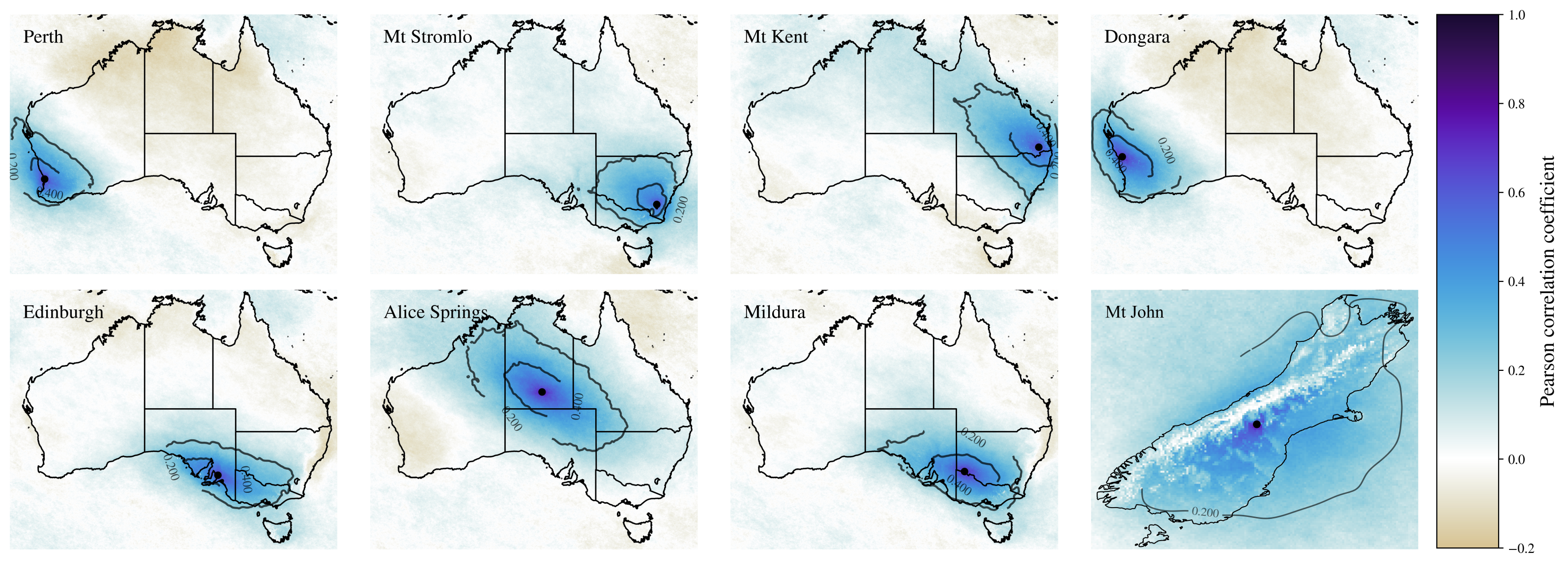}
  \caption{Surfaces of the Pearson correlation coefficient, $r_{ij}(\bm{\C}_k,\bm{\C}_{ij})$, between site $k$ in the network and all other pixels, $\bm{\C}_{ij} = \bm{\C}(x_i,y_j)$, for the Australasian region. Correlation surfaces are constructed by comparing the time-series of daily AHI/Himawari-8 cloud retrieval images between 2015-2022. We show contours for $r_{ij}(\bm{\C}_k,\bm{\C}_{ij})=0.2$ and $0.4$, overlaid on each surface.}
  \label{fig:ANZ_cloudcorr}
\end{figure*}

To provide a more robust estimate of $A$, we then calculate $A$ from all five satellites and decades of data for all sites of the extended AOGSN. All sources of remote sensing data outlined in \autoref{tab:sats} were used except for the Mt John OGS where the \textit{Aqua}/AIRS and MERRA-2 cloud retrieval algorithms produced erroneous results, likely due to nearby snow and ice or low resolution pixels including high altitude mountains. For low spatial resolution images, the closest pixel to the site is used to represent the sites cloud cover. Given Himawari-8's 500~m/pixel spatial resolution, a region-of-interest (ROI) is drawn around the sites and pixels within that ROI are averaged per frame. We model the spatial extent of these ROIs to correspond with a projected horizontal distance from a single layer of cloud at some height when observed at $\theta_\text{elevation}=30^\circ$\footnote{ $\theta_\text{elevation}=30^\circ$ is considered the minimum elevation angle for space-to-ground links throughout this paper.}. Himawari-8 is capable of detecting cloud height, so we measure 5~km as the mean cloud height over the entire Australasian region from 2015-2022, such that our ROI for Himawari-8 is approximately 9~km in radius, or $\pm2$~pixels.

\autoref{fig:aogsn_seasonal} shows $A$ across all satellites and the annual average of $A\pm\sigma$, with $\sigma$ from satellite and seasonal variation. From \autoref{fig:aogsn_seasonal} we see that most sites have high availability with an average of $A=0.69$, particularly the arid sites such as Alice Springs ($A=0.83$) and Dongara ($A=0.77$). Most sites exhibit a drop in $A$ for winter months, except for the northern sites of Alice Springs ($23.8^\circ S$) and Mt Kent ($27.8^\circ S)$, which experience a partial dry and wet seasonal cycle. Mt John is an outlier due to its low availability and pattern of clearer skies in the winter despite its southern latitude ($44^\circ$S). \autoref{sec:satellitevariation} demonstrates the independent results of $A$ from the different data sets used.

\section{Network diversity}
\label{sec:correlation}
\subsection{Spatial correlation of cloud cover}

Correlation of cloud cover between sites is the primary impact on network diversity, as two or more network nodes may be affected by the same weather. An effective FSOC network should be designed such that its OGS do not have strongly correlated cloud cover. For the case of GEO or deep space communication where visibility is maintained to an entire network, diversity corresponds directly to the likelihood that a link can be established with said network.

We use Himwari-8 for all cloud correlation analysis as it has made simultaneous $\C$ measurements of the entire region over the last eight years. Each pixel is flagged as $0$ if no cloud or only cirrus cloud is present, otherwise it is flagged as $1$, as we described in \autoref{sec:cover}. Therefore we have the vector $\bm{\C}_{ij}$\footnote{Throughout this study, we use the subscript $X_{ij}$ for function $X(x_i,y_j)$ evaluated at grid coordinates $(x_i,y_j)$.} of length $n$, where $n$ is the number of daily images from 2015-2022, for each $5$~km$\times5$~km pixel $(x_i,x_j)$ in Himawari-8 frames of Australasia. We calculate the point-wise Pearson (linear) correlation coefficient, $r_{ij}$, between site $k$, $\bm{\C}_k$, and every other pixel in Australasia, $\bm{\C}_{ij}$.
\begin{equation}
    \label{eq:pearsoncorrcoef}
    r_{ij}(\bm{\C}_k,\bm{\C}_{ij})=\frac{ \Exp{\Big(\bm{\C}_k-\Omega_k\Big)\Big(\bm{\C}_{ij}-\Omega_{ij} \Big)}_t }{ \left[\Big\langle \Big(\bm{\C}_k-\Omega_k\Big)^2\Big\rangle_t \Big\langle \Big(\bm{\C}_k-\Omega_{ij}) \Big)^2 \Big\rangle_t\right]^{1/2} },
\end{equation}
where $-1 \leq r_{ij} \leq 1$. For $r_{ij}=1$ the cloud cover between two pixels is perfectly linearly correlated\footnote{Note that this type of correlation only captures linear correlations, and more complex, non-linear correlations may exist but will not be accounted for with this statistic, see e.g. \cite{Tostheim2018_correlation}.} in time, hence when one pixel is cloudy, so is the other one. Conversely, when $r_{ij}=-1$ the pixels have perfect linear negative correlation, and when one pixel is cloudy the other is not. Finally, $r_{ij}=0$ corresponds to the case where two pixels are not linearly correlated in time. $r_{ij}(\bm{\C}_k,\bm{\C}_{ij})$ is therefore a two-dimensional surface describing how $\C$ throughout the region is spatially correlated around site $k$. The length scale and shape of $r_{ij}$ is driven by localised and global weather dynamics, e.g. atmospheric circulation cells on the global scale and rain shadow more locally.

\autoref{fig:ANZ_cloudcorr} shows $r_{ij}(\bm{\C}_k,\bm{\C}_{ij})$ for each of site in the extended AOGSN. We find contours of equal $r_{ij}$ around sites that represent the weather dynamics driving $\C$, e.g. elongation along the East-West axis for southern sites due to prevailing westerlies across Australia. From \autoref{fig:ANZ_cloudcorr}, we see that Alice Springs experiences correlated $\C$ over an extremely large area, the result of how pressure systems and fronts traverse the desert without coastal or topological effects. It is also evident that Dongara and Perth are located within each others $r_{ij}=0.4$ contour, implying they will be strongly correlated as we show in \autoref{sec:sitespecificcorr}. The Mt John site highlights the effect of rain shadow, as the correlation scale is stretched along the length of the New Zealand Southern Alps. Mt John correlations also show misleading results on the alps, due to those pixels sitting above cloud height. Negative correlation between northern Australia and south western Australia is also clear in \autoref{fig:ANZ_cloudcorr}, which occurs because northern Australia experiences has a tropical seasonality, i.e. a winter dry season, unlike the rest of the continent.

This method of computing $r_{ij}(\bm{\C}_k,\bm{\C}_{ij})$ for $k$ sites is an extension earlier work~\cite{fuchs2015ground}, i.e. we compute $r_{ij}$ against $\C_k$ for the whole grid $(x_i,y_j)$, and not just point-wise between preselected site locations. This spatially-resolved approach, despite being more computationally expensive, allows us to construct a two-dimensional surface of $r_{ij}$ for each site. In \autoref{sec:optimisation}, we show how this quantity can be used to generate a network of maximally diverse ground stations.

\subsection{Site-specific correlations}
\label{sec:sitespecificcorr}

\begin{figure}[!ht]
  \centering
  \includegraphics[width=\linewidth]{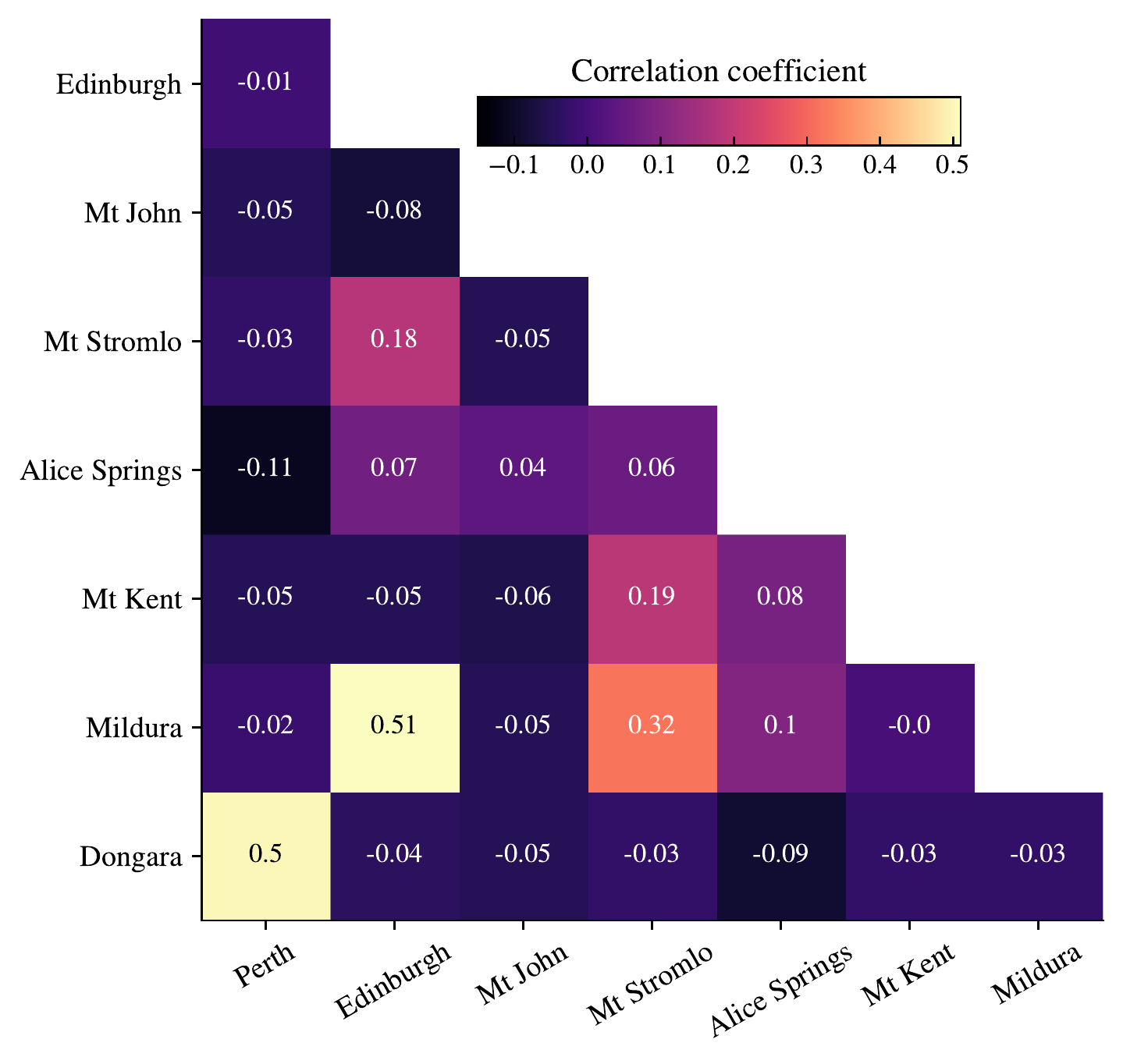}
  \caption{Pearson correlation matrix (\autoref{eq:pearsoncorrcoef} for each pair-wise site) for all sites in extended AOGSN calculated using daily cloud retrieval frames of Australasian region since 2015 from Himawari-8.}
  \label{fig:corr_matrix}
\end{figure}

To probe the pair-wise correlations between sites we calculate $r(\bm{\C}_k,\bm{\C}_l)$ between all sites of the extended AOGSN, where $k,l\in\{1,\dots,N\}$ for a network of $N$ sites. We show the results with a correlation matrix in \autoref{fig:corr_matrix}. \autoref{fig:corr_matrix} shows typically low, $r\lessapprox0.1$, between sites. However, we do note the Mildura-Edinburgh ($r=0.51$) and Perth-Dongara ($r=0.5$) sites show relatively strong, positive correlations. For example, this corresponds to half of all outages at Perth or Dongara affecting both sites. Negative correlation between Alice Springs and Perth/Dongara is due to the winter dry season of northern Australia, visible in \autoref{fig:ANZ_cloudcorr}.

As a simple metric to understand how diverse a network of $N$ sites is, we compute the mean absolute correlation in the network, 
\begin{equation}
    \rabs ={N\choose2}^{-1}\sum_{k=1}^{N-1}\sum_{l=k+1}^{N}\Big\vert r(\bm{\C}_k,\bm{\C}_l)\Big\vert,
    \label{eq:meanpearson}
\end{equation}
introduced earlier~\cite{fuchs2015ground}\footnote{Note, that compared to the value of $\bar{r}$ in \cite{fuchs2015ground}, we choose to take the mean of the absolute value, $\rabs$ which means we are probing the average magnitude of the correlation over the network as opposed to the average correlation. The latter may be zero for a network that has both strong positive and negative correlations, hence $\bar{r} = 0$ does not imply that the network is, as a whole, has weak correlations, but $\rabs=0$ does.}. We find $\rabs=0.088\pm0.11$ and $\rabs=0.073\pm0.076$ for the extended and base AOGSNs respectively, which is reasonably low despite being skewed by a number of strongly correlated sites (e.g., Mildura-Edinburgh) that were shown in \autoref{fig:corr_matrix}. Negatively correlated sites contribute given the absolute approach of \autoref{eq:meanpearson}, and the detrimental effects of negative correlations between network sites is explored in \autoref{sec:anticorroutages}. Nonetheless, $\rabs=0.088$ is representative of a well-diversified network, owing to the large area of the Australasian region. This highlights the unique position for Australia and New Zealand to offer a robust, highly-diverse FSOC ground segment, when combined with the high availabilities estimated in \autoref{sec:cover}.

\section{Network reliability and outage probability}
\label{sec:div}

The diversity and availability of a network can be used to estimate the network reliability and outage probability. To understand the outage probability, we construct $p_\text{avail}(k\leq M)$, the probability that at least $M$ sites in the network of $N$ nodes ($M \leq N$) are available in a network (the cumulative distribution function of the network availabilities.). The case of $M=0$ provides the network-wide outage probability, $p_\text{avail}(M=0)$. For GEOs and deep space targets, which can maintain visibility with an entire hemisphere $1-p_\text{avail}(M=0)$ corresponds directly to the network reliability. \autoref{fig:corr_matrix} shows that some significant correlations exist between AOGSN nodes, as do the networks proposed in \cite{fuchs2015ground}. Therefore, it is necessary to construct $p_\text{avail}(k\leq M)$ including point-wise correlations between sites.

Constructing a closed-form analytical expression for $p_\text{avail}(k\leq M)$ with correlation maintained between greater than three sites is impossible \cite{dai2013multivariate}. An earlier study provided an expression for uncorrelated sites and empirically measured the outage probabilities from $\bm{\C}$ data for the correlated case \cite{fuchs2015ground}. Empirically estimating $p_\text{avail}(k\leq M)$ can be compromised by the length of time required to determine low probability events such as a network-wide outages \cite{fuchs2015ground}. For our case, we use only twice daily images, and find only one instance of a network-wide outage over the 2015-2022 time span for the extended AOGSN. This means empirically, similar to \cite{fuchs2015ground}, we cannot provide a robust estimate of $p_\text{avail}(M=0)$ and require a better method to estimate these low probability, correlated events. A robust method can also be used to estimate from correlated weather data built up over a relatively short amount of time.

Because there is no closed-form analytical expression for the probability function of a correlated network of Bernoulli events, we opt for a Monte Carlo sampling approach. Our goal is to construct $p_\text{avail}(k\leq M)$ from a distribution that we are able to randomly draw events from. As per \autoref{sec:cover}, each event is a time realisation of the cloud cover at each site, $\C_k$. We therefore have to construct the the joint availability distribution (a multivariate Bernoulli distribution of dimension $N$) with correlations between each $\C_k$ marginal distribution. Again, $k,l\in\{1,\dots, N\}$ for a network of $N$ sites. We do this by utilising a neuronal spike model~\cite{Macke2009_neuronal_spike_model}, noting that a ground station network described by a correlated network of Bernoulli events is directly analogous to the correlated network of firing neurons~\cite{Macke2009_neuronal_spike_model}. 

Following \cite{Macke2009_neuronal_spike_model}, the method of constructing and sampling from the joint availability distribution is as follows. Consider $\bm{x} \in \mathbb{R}^{N}$, with components $x_k$. Let $\bm{x}$ follow a multivariate Gaussian distribution (that is, each $x_k,$ admits to a Gaussian distribution),
\begin{align}
    \label{eq:multi_gauss}
    f(\bm{x}; \bm{\mu}, \bm{\Lambda}) = (2\pi)^{-\frac{n}{2}}\text{det}(\bm{\Lambda})^{-\frac{1}{2}}\exp\left\{-\frac{1}{2}(\mathbf{x}-\bm{\mu})^{T}\bm{\Lambda}^{-1}(\bm{x}-\bm{\mu}) \right\},
\end{align}
with mean $\bm{\mu} \in \mathbb{R}^{N}$ and covariance $\bm{\Lambda} \in \mathbb{R}^{N\times N}$, a symmetric and positive-definite matrix and $\text{det}(\bm{\Lambda})$ is the matrix determinant. \cite{Macke2009_neuronal_spike_model} showed that if we are able to sample $f(\bm{x}; \bm{\mu}, \bm{\Lambda})$ and pick a threshold,
\begin{align}\label{eq:threshold}
    \C_k = 1,\;\; \text{if}\;\; x_k > 0,
\end{align}
then we are able to simulate a correlated multivariate Bernoulli process, with $\Omega_k$ determined by $\mu_k$ and the covariance $\Gamma_{kl}$ between each ground station determined by $\Lambda_{kl}$. Assuming that $\Lambda_{kk} = 1$, the relations between these quantities are as follows,
\begin{align}
    \Omega_k &= \phi(\mu_k), \\
    \Gamma_{kk} &= \phi(\mu_k)\phi(-\mu_k),\label{eq:gen_gamma_kk} \\
    \Gamma_{kl} &= \phi_{2}(\mu_k,\mu_l,\Lambda_{kl}) - \phi(\mu_k)\phi(\mu_l),\; k \neq l,
\end{align}
where
\begin{align} \label{eq:normal_map}
    \phi(\mu_k) = \frac{1}{2}\left[1 + \text{erf}(\mu_k/\sqrt{2})\right],
\end{align}
is the cumulative distribution function for a univariate, standardised ($\mu = 0$; $\Gamma=1$) Gaussian distribution, and $\phi_{2}(\mu_k,\mu_l,\Sigma_{kl})$ is the two-dimensional counterpart. $\mu_k$ is determined simply by inverting \autoref{eq:normal_map}, which can then be used to generate $\Gamma_{kk}$ using \autoref{eq:gen_gamma_kk}. For the off-diagonal elements of $\bm{\Lambda}$, we perform a root-finding algorithm to determine what values of $\Lambda_{kl}$ satisfy 
\begin{align} \label{eq:root_find}
    \Gamma_{kl} - \left[\phi_{2}(\mu_k,\mu_l,\Lambda_{kl}) - \phi(\mu_k)\phi(\mu_l)\right] = 0,
\end{align}
noting that $-1 \leq \Lambda_{kl} \leq 1$ (this is true for Bernoulli events, see Appendix~A in \cite{Macke2009_neuronal_spike_model}, but not generally true for other random processes) and that $\Gamma_{kl}$ is determined directly from the data (i.e., we compute the covariance between ground stations of the network, which from \autoref{eq:pearsoncorrcoef} one can see that it is $[\langle(\bm{\C}_k-\Omega_k)^2\rangle_t \langle(\bm{\C}_l-\Omega_{l}))^2 \rangle_t]^{1/2} r(\bm{\C}_k,\bm{\C}_l)$. $\Gamma_{kl}$ and $\Lambda_{kl}$ are monotonic functions of one another, and therefore \autoref{eq:root_find} has a single root between $-1$ and $1$, which we determine using \textsc{scipy} \cite{virtanen2020scipy}. This method differs from \cite{net2016approximation} by numerically solving \autoref{eq:root_find} instead of obtaining an analytical expression by approximation.

\begin{figure}[!ht]
  \centering
  \includegraphics[width=\linewidth]{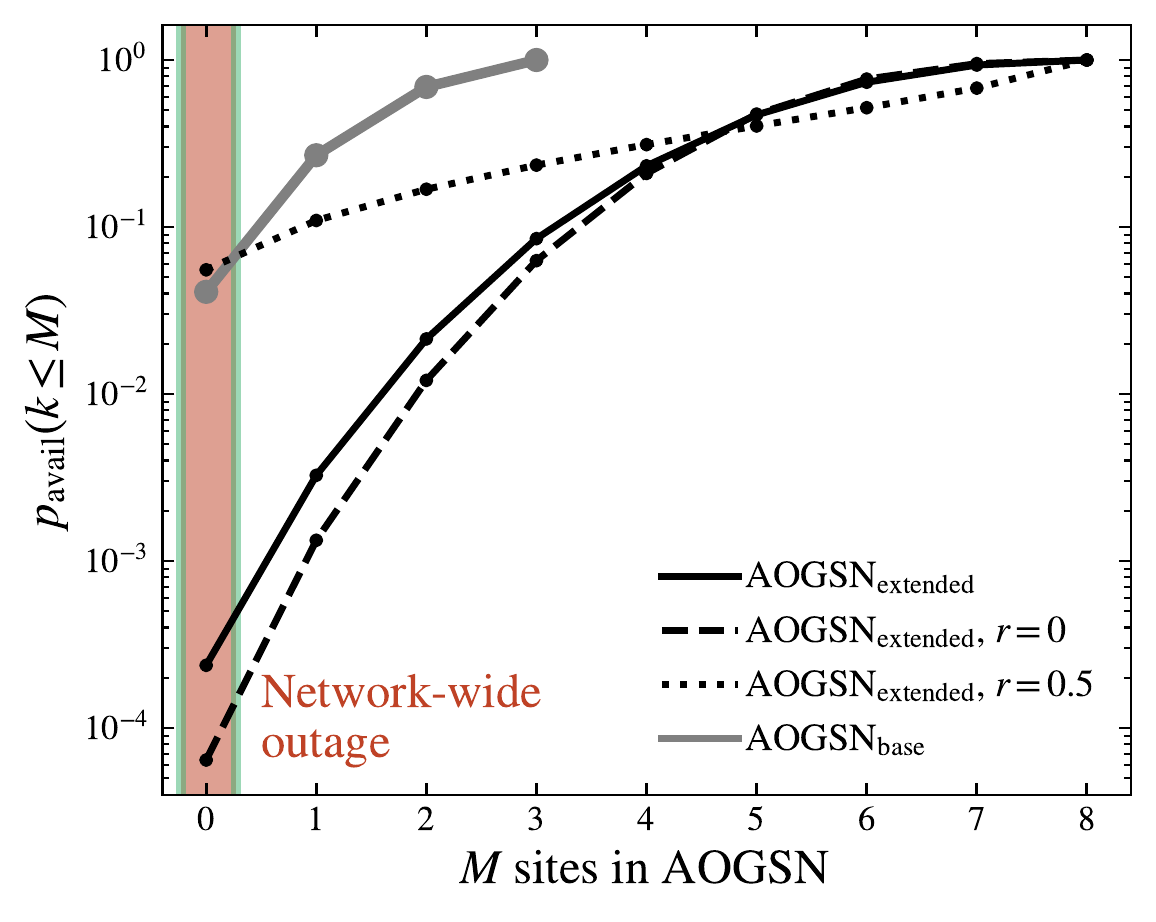}
  \caption{Cumulative distribution functions $p_{\rm avail}(k\leq M)$, for $k\leq M$ available sites, constructed using our Monte Carlo method outlined in \autoref{sec:div} for the extended (solid black line; $M=8$) and base (solid grey line; $M=3$) AOGSN, including point-wise correlations between ground stations (see \autoref{sec:div}). We show uncorrelated ($r=0$; black dashed line) and highly correlated ($r=0.5$; black dotted line) network availability results to compare with the extended and base AOGSN. We have highlighted the special case of $p_\text{avail}(M=0)$ (red and green band) representing a network-wide outage.}
  \label{fig:outageprobcurves}
\end{figure}

Utilising this construction and sampling method, we create three multivariate availability distributions for the extended AOGSN, and three availability cumulative distribution functions by sampling the availability distribution $10^8$ times (equivalent to a $10^8$ time realisations), all which use $\Omega_i$ directly determined from the data. The difference between the three distributions is as follows: (1) we construct a distribution with $\bm{\Gamma}$ directly from our data, encoding the real $r(\C_k,\C_l)$ between sites; (2) we run a $r(\C_k,\C_l) = 0, \; k \neq l$ simulation, which enforces that there is no correlation between sites, comparable to the results in \cite{fuchs2015ground}; and (3) we choose a $r(\C_k,\C_l)=0.5, \; k \neq l$ to understand the outage probability for a network that is strongly correlated in $\C$. 

\autoref{fig:outageprobcurves} shows the cumulative distribution function $p_\text{avail}(k\leq M)$ for $M=0-8$, with cases (1), (2), and (3) for the extended AOGSN, and with only case (1) for the base AOGSN. We estimate network-wide outage probabilities of $p_\text{avail}(M=0)=2.4\times10^{-4}$ and $p_\text{avail}(M=0)=0.064$ for the extended and base AOGSN configurations respectively. From \autoref{fig:outageprobcurves} we see that a small ($N=3$) and well-separated network, e.g., the base AOGSN, can have (marginally) lower or comparable outage probability than a highly available but very correlated network of eight OGS. \autoref{fig:outageprobcurves} also shows while $p_\text{avail}(M=0)=2.4\times10^{-4}$ for measured $r(\C_k,\C_l)$, $p_\text{avail}(M=0)=6.1\times10^{-5}$ for case (2), i.e., uncorrelated sites with $r(\C_k,\C_l)=0$. An analytical expression exists for $r=0$~\cite{fuchs2015ground}, but we can see that realistically modelling correlations results in a $4\times$ higher probability of a network-wide outage. Furthermore, for more correlated and dense networks, e.g. the Japanese network outlined in \cite{toyoshima2014introduction}. This discrepancy can be orders of magnitude as we see with case (3) for $r(\C_k,\C_l)=0.5$. \autoref{sec:anticorroutages} shows a negatively correlated network, $r(\C_k,\C_l)<0$, demonstrating how any correlations (negative or positive) increase $p_\text{avail}(k\leq M)$ compared to a $r(\C_k,\C_l)=0.0$ network, in agreement with \cite{net2016approximation}. 

This method shows how robust estimates can be produced for the network outages of a correlated network using availability and correlation data. It also highlights how a sparse, highly available, and diverse network, such as the example network we study, can provide an extremely reliable ground segment, overcoming one of the most significant barriers for space-to-ground FSOC. Next, we consider the case where we want to generate our own optimal network that maximises availability and diversity.

\section{Generating an optimal network}
\label{sec:optimisation}

We are interested in developing a method for site selection in an FSOC network. An optimised network, produced without preselection, can also be compared with an existing network, e.g. our example AOGSN, to judge how well-optimised the network is. This method could also be adopted to select additional sites in an established network. Methods have been presented for how diversity can be optimised with varying channel and outage models, but their optimisation relies on preselected sites in a network~\cite{erdogan2021site,lyras2019medium}.

\autoref{fig:ANZ_cloudcorr} indicates that correlation surfaces can be used for optimally planning an FSOC network. We expand this idea by considering a simple spatially-resolved model for iteratively determining the coordinates $(x_i,y_j)$ of $N$ sites $\mathcal{S}_k$, in the network $\N$ (the same notation as we used in \autoref{sec:cover}). Our method for determining the optimal $(x_i,y_j)$ for each $\mathcal{S}_k$ in the network relies upon two site characteristics: the first is average cloud cover $0\leq\Omega_{k}\leq 1$ at each $\mathcal{S}_k$. The second is the temporal correlation $-1\leq r_{ij}(\bm{\C}_k,\bm{\C}_{ij})\leq 1$. We discuss the exact procedure for generating the correlation in \autoref{sec:correlation}. To maximise the connectivity of each $\mathcal{S}_k$ and robustness of $\N$, we propose a simple heuristic: (1) the optimal site is one that minimises the amount of cloud coverage, hence maximising the time it may be available on the $\N$, and (2), the optimal $\N$ is one that supports a series of uncorrelated $\mathcal{S}_k$ (supported by the lowest probability of an outage in \autoref{fig:outageprobcurves}), such that if any $\mathcal{S}_k$ are unavailable due to cloud cover events, other sites in $\N$ will most likely not suffer the same deficiency. 

Based on this simple heuristic, the first site $\mathcal{S}_1$ we pick is located at,
\begin{align}
    \label{eq:minimise_S1}
    \mathcal{S}_1(x_i,y_j) = \text{argmin}_{ij}\left\{\Omega_{ij}\right\},
\end{align}
where $\text{argmin}_{ij}$ returns the sites coordinates that minimise the time-average of $\bm{\C}_{ij}$, $\Omega_{ij}$, without any extra constraints. The second site we pick does not, however, have the same degrees of freedom, and we want to minimise both $\Omega_{ij}$ and the site-pair correlation with $\mathcal{S}_1$. Hence, we define a new, quadratic (positive, semi-definite) function, $g$, as a function of spatial coordinates $(x_i,y_j)$ such that
\begin{align}\label{eq:objective_function}
    g_{ij} = \Big\|\omega_0\Omega_{ij} + \omega_1r_{ij}(\bm{\C}_1,\bm{\C}_{ij})\Big\|^2,
\end{align}
where $\omega_0$ and $\omega_1$ are weights that determine what we deem more important -- the cloud cover between sites, or maintaining an uncorrelated network, and $\|a + b\|^2 = a^2 + b^2$ is the two-norm squared. Note that $g_{ij} = 0$ at its global minima, which corresponds to a site with no average cloud cover, and no correlation with $\mathcal{S}_1$, and $g_{ij} = \omega_1^2+\omega_2^2$ at its maxima, describing a site that is on average always cloudy, and strongly correlated (either positive or negative\footnote{Note that we show that both types of correlations reduce the performance of the network in \autoref{sec:anticorroutages}}) with $\mathcal{S}_1$. Hence, $\mathcal{S}_2$ is
\begin{align} 
    \label{eq:minimise_S2}
    \mathcal{S}_2(x_i,y_j) = \text{argmin}_{ij}\left\{g_{ij}\right\}.
\end{align}

Now we must generalise \autoref{eq:objective_function} for a general $\mathcal{S}_k(x_i,y_j)$, noting that we ought to be always able to minimise $g_{ij}$ to find any next site on the network. The procedure for $N$ sites is simply,

\begin{align}
    \label{eq:objective_function_2}
    g_{ij}(N) =& \Big\|\omega_0\Omega_{ij} + \frac{1}{N}\sum_{k=1}^N\omega_kr_{ij}(\bm{\C}_k,\bm{\C}_{ij})\Big\|^2,\\
    =& \Big\|\omega_0\Omega_{ij} + \frac{1}{N}\left[\omega_1r_{ij}(\bm{\C}_1,\C_{ij})\right. +\nonumber \\
    & \left. \omega_2r_{ij}(\bm{\C}_2,\bm{\C}_{ij}) + \hdots + 
    \omega_Nr_{ij}(\bm{\C}_N,\bm{\C}_{ij})\right]\Big\|^2,
\end{align}

which describes our selection criteria, $g_{ij}(N)$ as a linear combination of time-averaged cloud cover through the first term, and pair-wise site correlations, each with a weighting $\left\{\omega_1,\hdots, \omega_N \right\}$. This maintains $g_{ij}(N) = 0$ as a global minima, but now $g_{ij}(N)=\omega_0^2 + (\omega_1^2+\omega_2^2+\hdots+\omega_N^2)/N^2$ is the global maximum. Hence, to generate the first site in $\N$ we use \autoref{eq:minimise_S1}, and then to successively generate $n\in N$ following sites we use \autoref{eq:objective_function_2}, i.e., 
\begin{align} \label{eq:minimise_SN}
    \mathcal{S}_n(x_i,y_j) =  \text{argmin}_{i,j}
    \begin{cases}
    \left\{\Omega_{ij}\right\}, & \text{ if } n=1, \\
    \left\{g_{ij}(n)\right\}, &  \text{ if } n>1,
    \end{cases}
\end{align}
to populate $\N$. Note that this simple method can be easily generalised for including any other site characteristics that may help determine a robust $\N$. This method can also be used to optimally add OGS to an existing network, explored in \autoref{sec:roimasking}.

\begin{figure}[!ht]
  \centering
  \includegraphics[width=\linewidth]{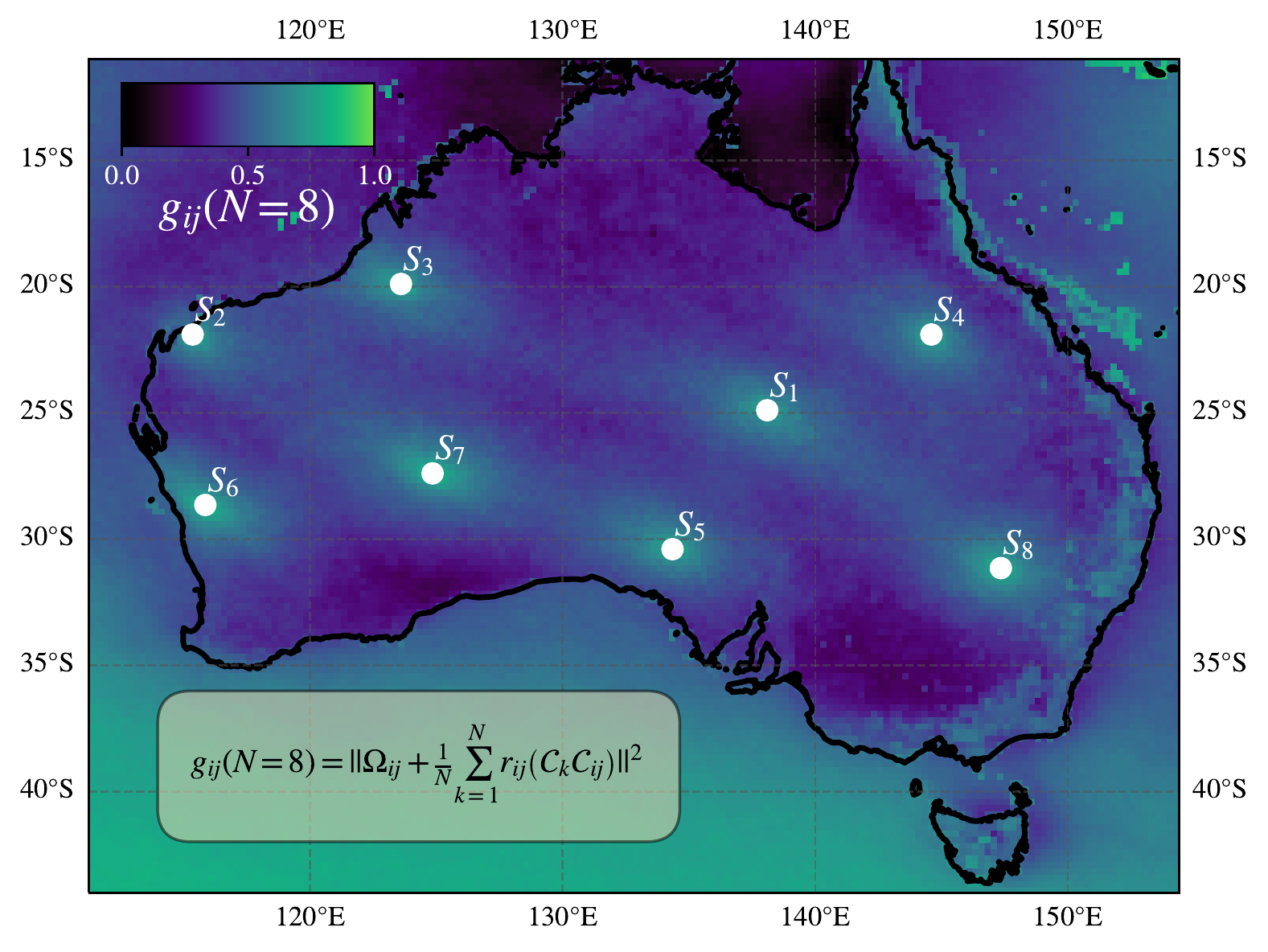}
  \caption{Distribution of $g_{ij}(N=8)$ function, \autoref{eq:objective_function}, and the minimum for linear combinations of availability and spatial correlation from Himawari-8 cloud retrieval data. Eight optimally selected sites, $\N_{N=8}$, are highlighted with white dots and indexed from $k=1$.}
  \label{fig:beattiemap}
\end{figure}

We initially set all $\omega=1$, however \autoref{sec:networkcapacity} of \autoref{sec:coverage} explores the use of a $\omega$ function to further optimise for a given orbital inclination. The $\omega$ terms can also be used to mask for regions-of-interest. The result of running this method with eight years of daily cloud data from Himawari-8 over Australia is shown in \autoref{fig:beattiemap} for $N=8$. New Zealand is ignored for simplicity in this analysis and to reduce computational demand given the number of correlation calculations. 

We also reduce the spatial resolution of Himawari-8 data for this analysis to 25~km from 5~km to further reduce computational effort. \autoref{fig:beattiemap} is dark where $g_{ij}$ is close to $0$, and the process of minimising $g_{ij}$ selects these areas. Each $\mathcal{S}_k$ also imposes a structure to the surface of $g_{ij}$ related to the $r_{ij}(\C_k,\C_{ij})$ functions in \autoref{fig:ANZ_cloudcorr}. We see from \autoref{fig:beattiemap} that sequentially minimising $g_{ij}(N)$ picks locations for $\N_{N=8}$ which are well-spaced and in areas of low cloud cover, tending to be in extremely arid parts of Australia. 

The effectiveness of our optimisation technique can then be judged by comparing with the extended AOGSN, a more realistic example network. \autoref{tab:networkcomp} uses the mean of $A$ across each of the sites, $\bar{A}$, $\rabs$ (\autoref{eq:meanpearson}), and $p_\text{avail}(M=0)$ to compare the AOGSN with $\N_{N=8}$. \autoref{tab:networkcomp} shows that for all metrics $\N_{N=8}$ outperforms the extended AOGSN. We also see in \autoref{tab:networkcomp} that $\N_{N=8}$ has an order of magnitude lower chance of a network-wide outage. However, the metrics we employ in \autoref{tab:networkcomp} do not properly represent network capability upon comparison, e.g. outage probability for both $\N_{N=8}$ and the extended AOGSN seem sufficient for a reliable network. In the next sections we use orbital simulations to provide a more detailed view of network capability.

\begin{table}[!ht]
    \centering
        \caption{Comparison between base and full AOGSN with optimised network, $\N$.}
            \begin{tabular}{p{2cm} | p{1cm} | p{1cm} | p{2cm}}
                \toprule
                Network & $\bar{A}$ & $\rabs$ & $p_\text{avail}(M=0)$\\
                \hline\hline
                AOGSN \:\:Extended & $0.69\pm0.09$ & $0.08\pm0.11$ & $2.4\times10^{-4}$\\
                \hline
                AOGSN Base & $0.67\pm0.06$ & $0.073\pm0.076$ & $0.064$\\
                \hline
                $\N_{N=8}$ & $0.83\pm0.04$ & $0.069\pm0.060$ & $1.47\times10^{-5}$\\
                \hline\hline
            \end{tabular}
    \begin{tablenotes}
        \item\textit{\textbf{Notes.}} $\bar{A}$ is statistical mean of $A$ across all sites in a given network and $\rabs$ is determined from \autoref{eq:meanpearson} shown with $\pm1\sigma$ from both calculations. $p_\text{avail}(M=0)$ is from the Monte Carlo method for sampling the availability PDF, described in \autoref{sec:div} The uncertainties of $A$, as with Fig \ref{fig:aogsn_seasonal}, represent seasonal variation and satellite-to-satellite variability. Uncertainties of $\rabs$ is the $\sigma$ from all values in the matrix used to compute $\rabs$ described in Equation \ref{eq:meanpearson}.
    \end{tablenotes}
    \label{tab:networkcomp}
\end{table}

\section{LEO satellite link modelling}\label{sec:coverage}

We conduct orbital propagation simulations that can be combined with availability measurements to estimate the data capacity for different network configurations. We do this by using the \textit{a.i.-solutions} FreeFlyer software to simulate LEO satellite links to the AOGSN. Coverage as a function of orbital height ($h_\text{orbit}$) and minimum OGS elevation angle ($\theta_\text{elevation}$) can be used to estimate the number and duration of passes a satellite has with the network. While GEO or deep space applications are likely to maintain mutual visibility across the entire network, LEO satellites will typically only see one node at a time, so an increase in number of passes directly corresponds to an increased number of links and average data capacity. 

\begin{figure}
  \centering
  \includegraphics[width=\linewidth]{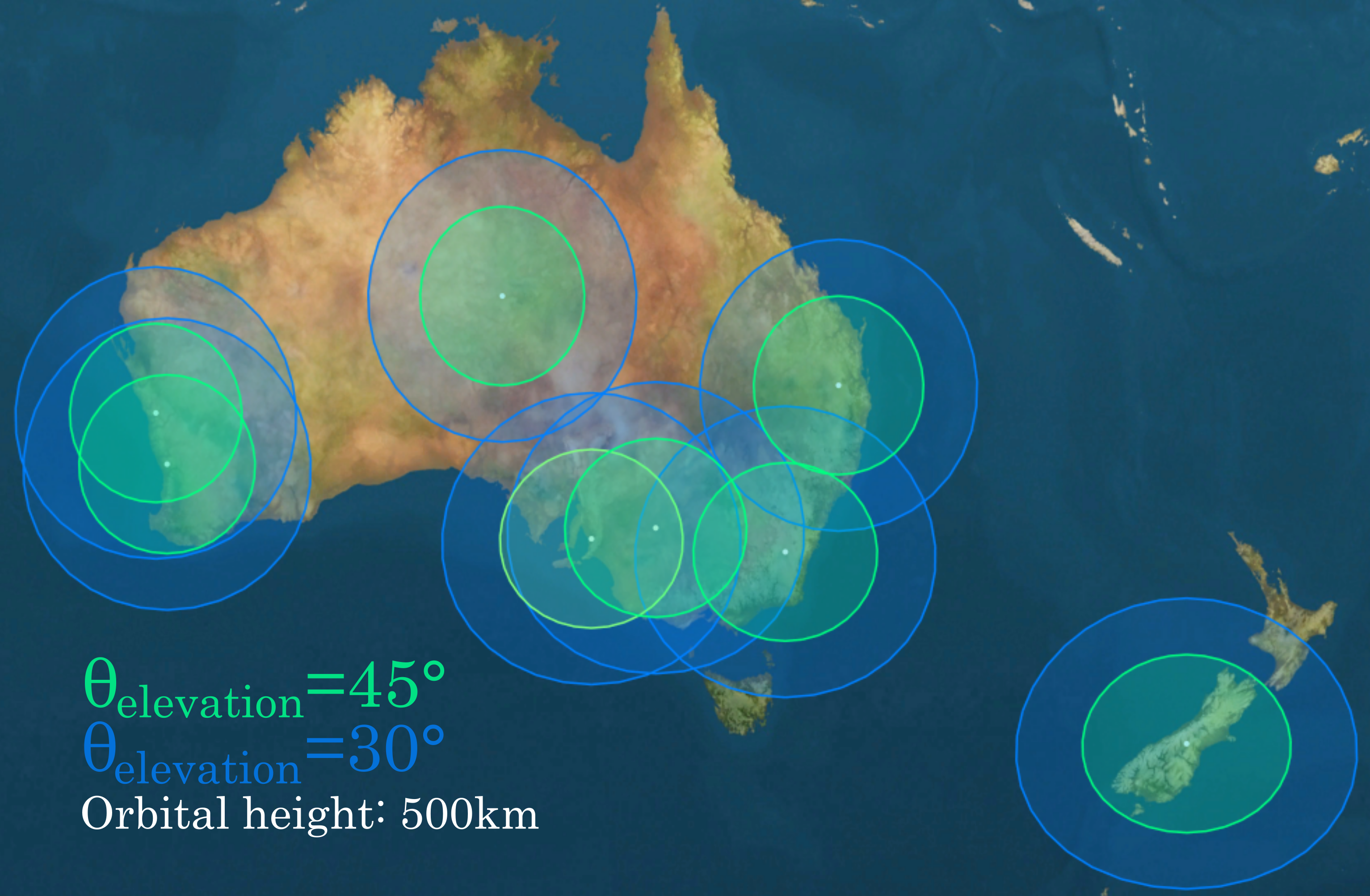}
  \caption{Coverage for the ground stations in the extended AOGSN with $\theta_\text{elevation}>30^\circ$ (blue) and $\theta_\text{elevation}>45^\circ$ (green)  to satellite with $h_\text{orbit}=500$~km.}
  \label{fig:coveragefigure}
\end{figure}

\autoref{fig:coveragefigure} shows coverage for ground stations in the AOGSN with $\theta_\text{elevation}\geq30^\circ$ and $\theta_\text{elevation}\geq45^\circ$ to a satellite with $h_\text{orbit}=500$~km. $h_\text{orbit}=500$~km is selected as a typical orbit for LEO satellites. The area covered by the network in \autoref{fig:coveragefigure} is large, highlighting how an Australasian network can provide ground segment support over a significant area, continuing a well-established role in radio ground segment operations~\cite{del2016architecting}. However, the overlapping circles in \autoref{fig:coveragefigure} represent areas of mutual coverage, implying the proposed extended AOGSN has some redundancy not present in the optimised network, $\N_{N=8}$.

\subsection{Link duration analysis}

\begin{figure}[!ht]
  \centering
  \includegraphics[width=\linewidth]{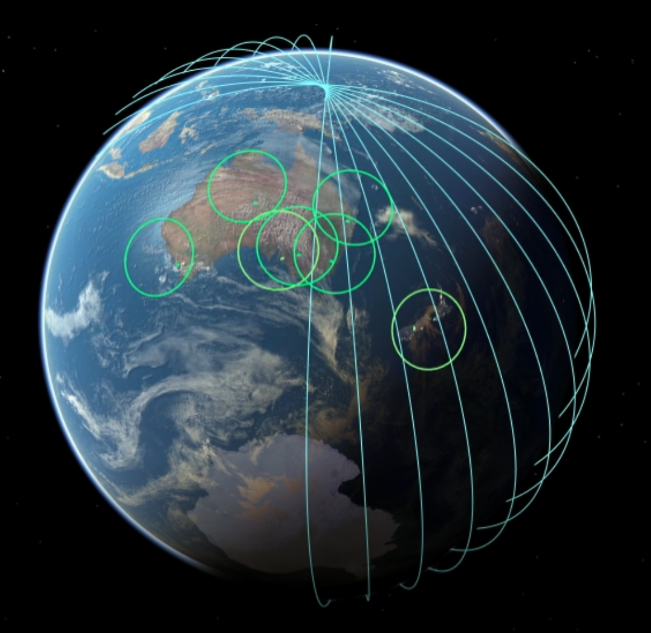}
  \caption{A single time-step from \textit{a.i.-solutions} FreeFlyer orbital simulations of AOGSN and satellites with inclination $i=10^\circ-100^\circ$ and $h_\text{orbit}=530$~km. Blue traces are recent satellite orbital trajectories. Green circles correspond to coverage for $\theta_\text{elevation}\geq30^\circ$. Note that the Dongara site is not visible in this sample.}
  \label{fig:dynamic_snapshot}
\end{figure}

We conduct further orbital simulations to determine average link durations $\tau(i)$ between LEO orbits and the extended AOGSN, i.e., the sum of passes and their duration. The parameter space of orbital inclination $i$ is modelled by scaling a fiducial ISS orbit to $530$~km (common height for LEO communication satellites), and then orientating it from $i=20^\circ-100^\circ$. These orbits are propagated for one year to create a robust statistical sample. This method may overestimate $\tau$ because it includes short passes that glance $\theta_{elevation}=30^\circ$ which would typically be ignored, and passes close to the Sun were not discarded. A single time step of this simulation is visible in \autoref{fig:dynamic_snapshot}, where sample orbits are shown in blue, and site coverage at $i=30^\circ$ in green. For each site in the AOGSN, all passes and their durations are totalled to determine $\tau(i)$ for $530$~km orbital height and a minimum $\theta_{elevation}=30^\circ$.

\begin{figure}
  \centering
  \includegraphics[width=\linewidth]{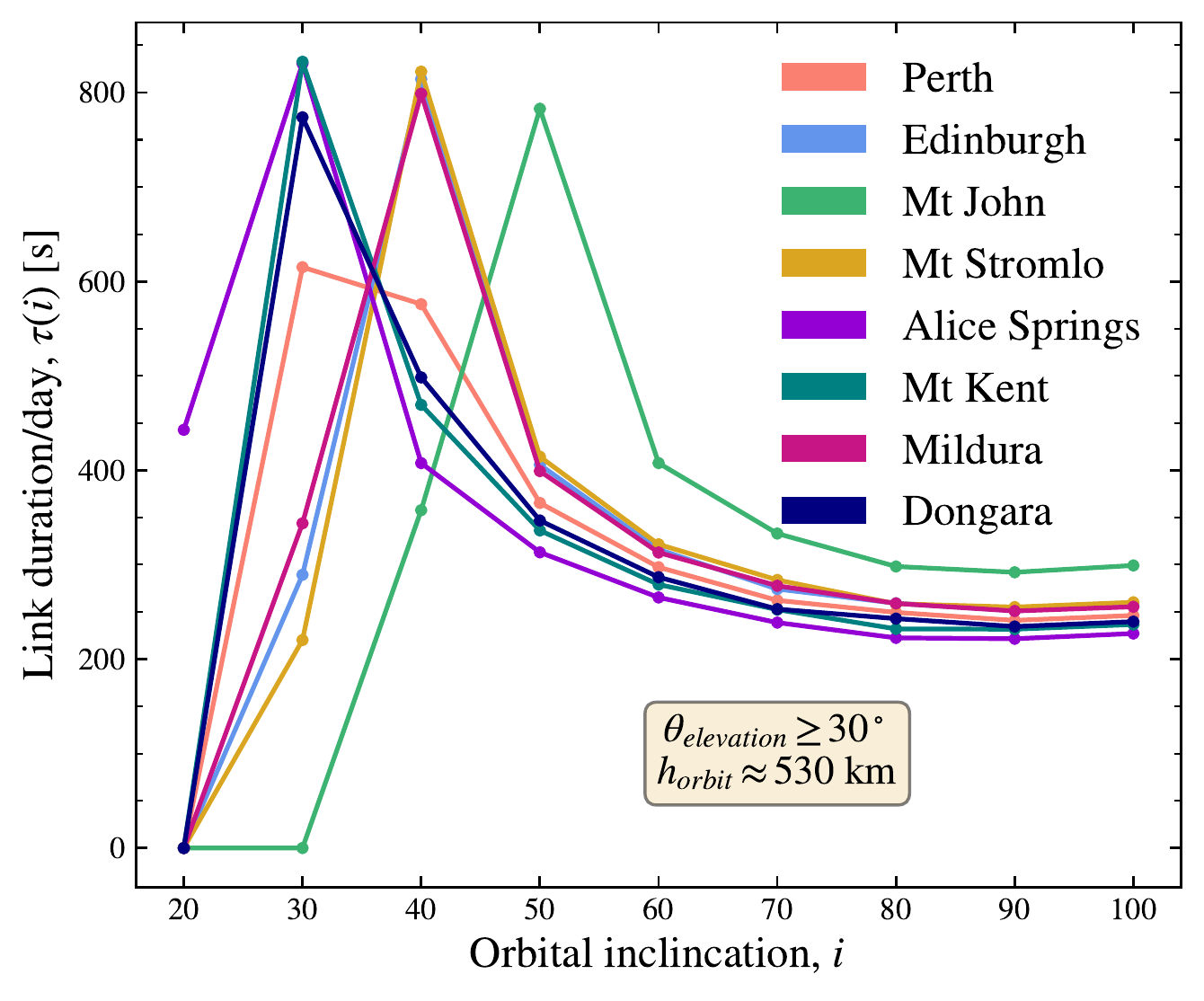}
  \caption{The daily average link duration, $\tau(i)$ (number of seconds/day) between each ground station (different colours) of the extended AOGSN and satellites with orbital inclinations from $20^\circ-100^\circ$. Availability via the cloud cover (see \autoref{sec:cover}) is not considered for these estimates. Orbital parameters and results are based on a fiducial ISS orbit scaled to $530$~km and propagated for one year.}
  \label{fig:link_durations}
\end{figure}

\autoref{fig:link_durations} shows the per day average of $\tau(i)$ from the year-long orbital simulation results. Each site has a similar curve, with a peak around a sites respective latitude, $L$. This relationship $\max\{\tau(i)\}\propto L$ demonstrates that networks and OGS placement can be optimised for a particular orbit or a given constellation, e.g. southern latitude (for the southern hemisphere) sites are preferable to link with polar orbiting satellites. 

\subsection{Latitude correction for network optimisation}
\label{sec:latitudecorr}

\autoref{fig:link_durations} shows the strong relation between site latitude, $L$, and $\tau(i)$. A network optimised for availability and diversity preferences the arid northern portion of Australia, such as $\N_{N=8}$ from \autoref{sec:optimisation}, which may not be ideal for maximising $\tau$ with a given satellite or constellation. We therefore adjust the method from \autoref{sec:optimisation} to produce an optimised network for a specific $i=90^\circ$ polar orbit satellite. Many current FSOC satellites are approximately in polar orbits, and Earth observation satellites, which may one day be supported by an FSOC ground segment for improved downlinks, typically use a Sun-synchronous near-polar orbit. 

To optimise for an $i=90^\circ$ orbit we look at the $\tau(i=90^\circ)$ values for each AOGSN site in \autoref{fig:link_durations}, and deduce a relation empirically from our simulations. If $L$ is far from $i$, we can see that $\tau(i)$ is asymptotic, which is satisfied for all sites in the extended AOGSN. We therefore record $\tau(i=90^\circ)$ as a function of $L$ and normalise the results,
\begin{align}
    \label{eq:lat_weighting_func}
    \omega(L) = 0.00745L+1
\end{align}
to create an expression that describes $\tau(i=90^\circ)\propto L$. 

\begin{figure}[!ht]
  \centering
  \includegraphics[width=\linewidth]{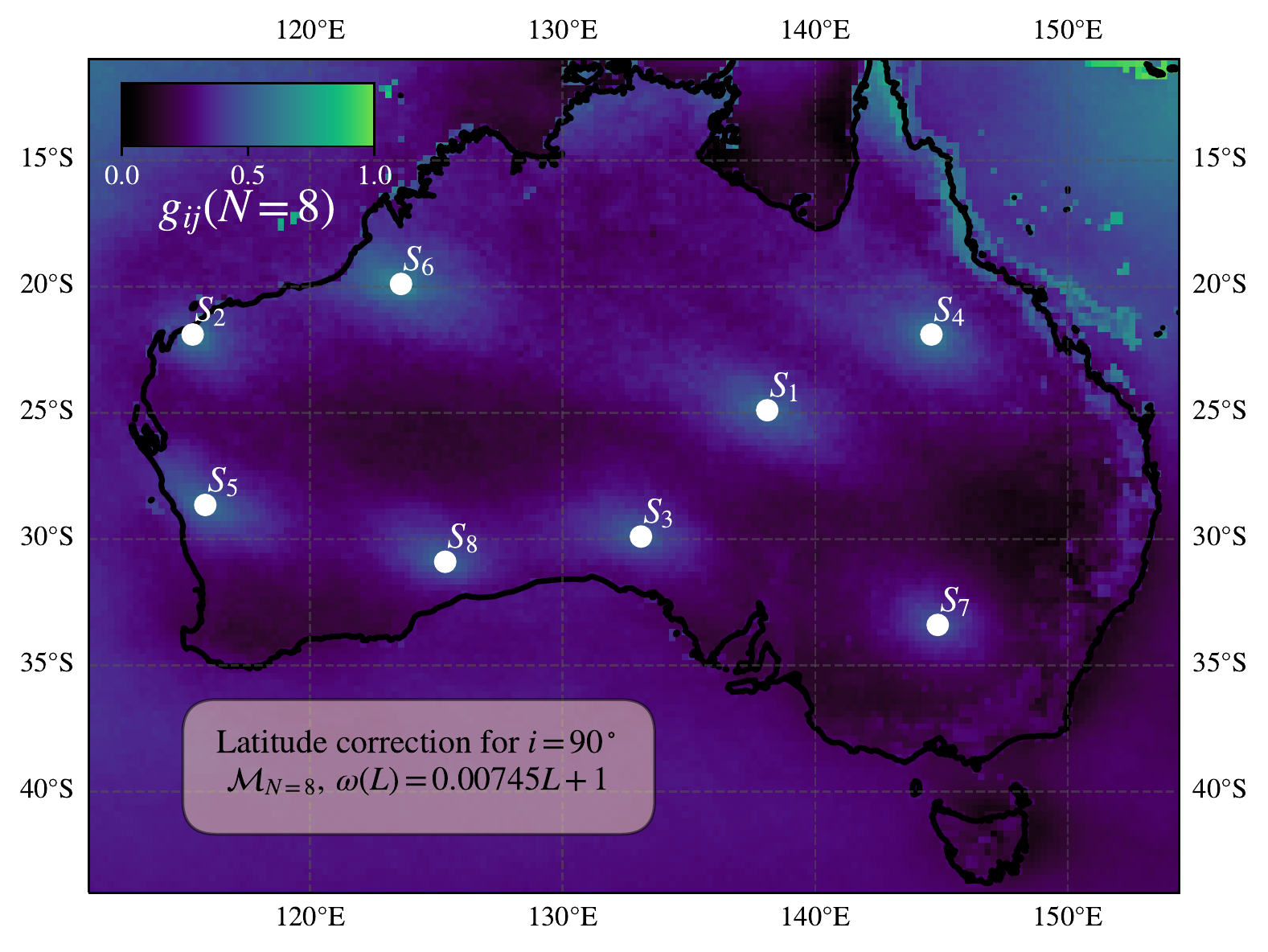}
  \caption{Optimised selection of eight sites, $\mathcal{M}_{N=8}$, using method from \autoref{sec:optimisation} with linear latitude weighting function, \autoref{eq:lat_weighting_func}, to optimise specifically for $i=90^\circ$ orbits.}
  \label{fig:latitudecorrectedmap}
\end{figure}

$\omega(L)$ is normalised to the equator ($L=0$) with a minimum of $0.33$ at the south pole ($L=-90$). We can then use $\omega(L)$ as a weighting function for all $\omega_k$ in \autoref{eq:objective_function}, which adds a linear North-South gradient for preferencing southern sites to the surface, $g_{ij}(N)$, being minimised by the algorithm in \autoref{sec:optimisation}. We re-run the optimisation to generate $\mathcal{M}_{N=8}$, the set of sites selected after applying this latitude-correction which is shown in \autoref{fig:latitudecorrectedmap}. We see very little difference between \autoref{fig:latitudecorrectedmap} ($\mathcal{M}_{N=8}$) and \autoref{fig:beattiemap} ($\N_{N=8}$) for $S_1$ to $S_6$, but with some southern shift for $S_7$ and $S_8$ due to the new correction term, $\omega(L)$. This similarity is likely because the first sites have such high availability that they remain optimal despite the orbital dynamics favouring southern locations. The method we have demonstrated could be generalised to any orbital parameters, not just $i$, and for an entire satellite constellation.

\subsection{Network capacity}
\label{sec:networkcapacity}

We finally combine work on availability, $\tau(i)$ modelling, and optimisation techniques by constructing a general network capacity for LEO satellites. The availability-corrected, network-wide link duration, $T(i)$, for a network of $N$ sites is,
\begin{equation} \label{eq:networkcapacityeq}
    T(i) = \sum_{k=1}^{N}A_k\tau_k(i).
\end{equation}

\begin{figure}[!ht]
  \centering
  \includegraphics[width=\linewidth]{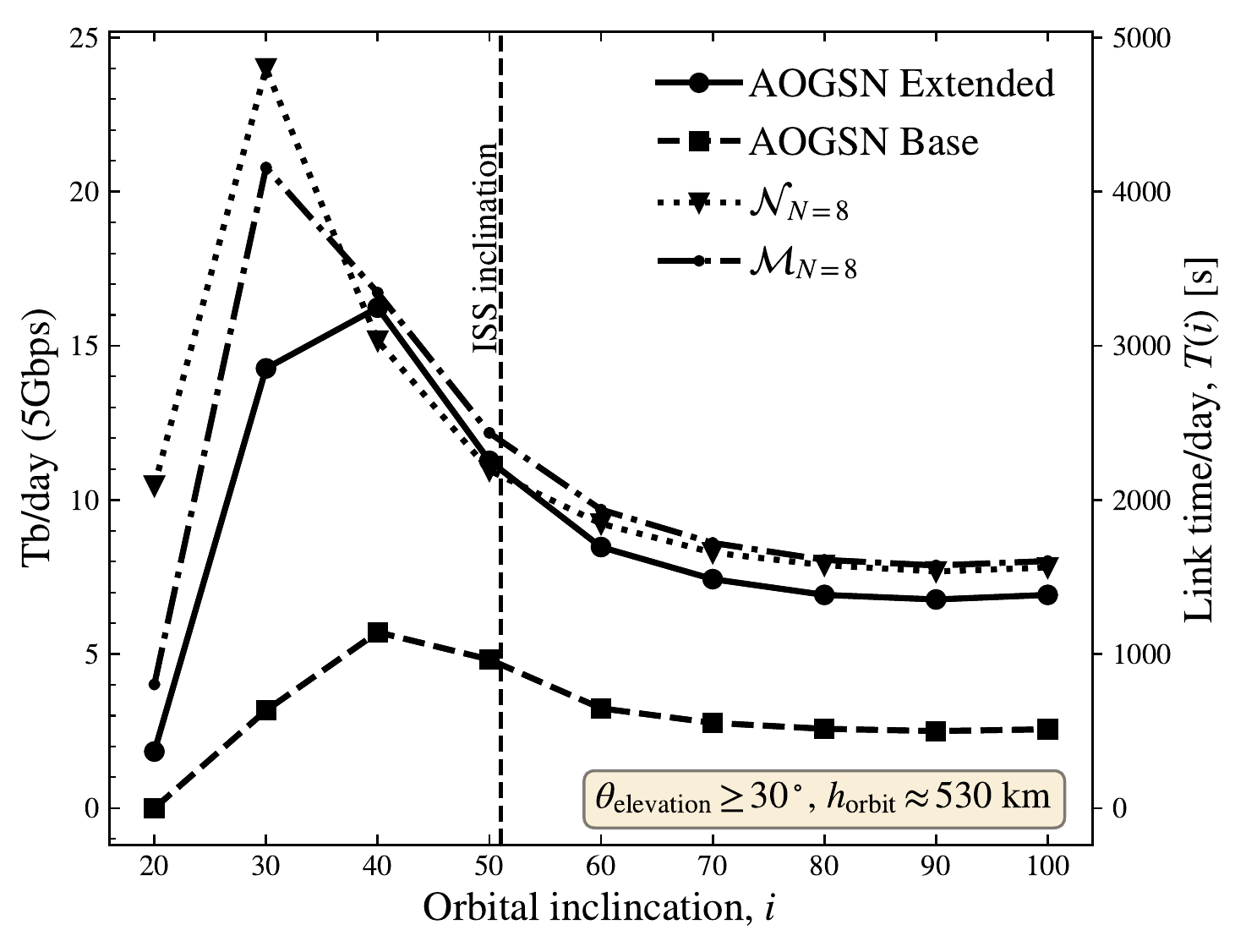}
  \caption{Network link duration, $T(i)$ (\autoref{eq:networkcapacityeq}), and data capacity estimates for the AOGSN (extended, solid; base, dashed) and two networks optimised for maximising availability and diversity. Optimised networks $\mathcal{M}_{N=8}$ (dashed-dotted, \autoref{sec:latitudecorr}) and $\N_{N=8}$ (dotted with triangles, \autoref{sec:optimisation}) are with and without latitude-correction for polar orbits respectively. Data estimation uses $5$~Gbps as the fiducial bitrate, multiplied by $T(i)$ for current on-off keying capability, assuming a uniform atmospheric channel for every site.}
  \label{fig:cloud_bitrate}
\end{figure}

We calculate $T(i)$ for four networks: (1) the base AOGSN, (2) extended AOGSN, (3) $\N_{N=8}$, and (4) $\mathcal{M}_{N=8}$. \autoref{fig:cloud_bitrate} shows $T(i)$ for these four networks, and an estimate of the daily data capacity with a $5$~Gbps bit rate, possible with future on-off keying satellite terminals \cite{schmidt2022dlr,carrasco2022development}. Note that this expression for $T(i)$ overestimates total link time by including scenarios where a LEO can maintain mutual visibility with $>1$ available sites, by including both links. In practice, the satellite must choose which OGS to link with, and the chance of both being available simultaneously is related to $r$ for that pair, e.g. from \autoref{fig:corr_matrix}. Estimating network capacity in this manner does not consider potential outages or degradation of bit-rate due to turbulence, so it is only a guide of system performance, for a given bit-rate.

\autoref{fig:cloud_bitrate} shows that both the AOGSN and algorithm-optimised networks can feasibly support many terabits of data transfer per day with a LEO satellite and current capability~\cite{carrasco2022development,schmidt2022dlr}. An general improvement can be seen in \autoref{fig:cloud_bitrate} between the extended AOGSN and optimised networks, $\N_{N-8}$ and $\mathcal{M}_{N=8}$. $\N_{N=8}$ and $\mathcal{M}_{N=8}$ have a $13\%$ and $16\%$ increased data capacity compared with the extended AOGSN respectively for $i=100^\circ$, approximately a Sun-synchronous polar orbit.  Therefore, the minor difference of $S_7,S_8$ $\in\mathcal{M}_{N=8}$, optimised with the $\omega(L)$ weighting function, shows only a marginal improvement of $3\%$ over $\N_{N=8}$ for highly inclined orbits.

We may be interested in the case of network capacity for a LEO constellation uniformly distributed in $i$. Integrating $T(i)$ over $i$ for all networks in \autoref{fig:cloud_bitrate} allows to compare this general case. Upgrading from the $N=3$ base AOGSN to $N=8$ extended AOGSN corresponds to a $293\%$ increase of data capacity when integrating $T(i)$ for the uniform case. This also shows a $27\%$ and $20\%$ efficiency gain for $\N_{N=8}$ and $\mathcal{M}_{N=8}$ respectively, compared to the extended AOGSN. The optimisation from \autoref{sec:optimisation} therefore yields a network that is superior in data capacity to the more realistic AOGSN. We could also use this method to optimally select additional sites for a network and estimate the increased capacity from doing so.

\begin{figure*}[!ht]
  \centering
  \includegraphics[width=0.8\textwidth]{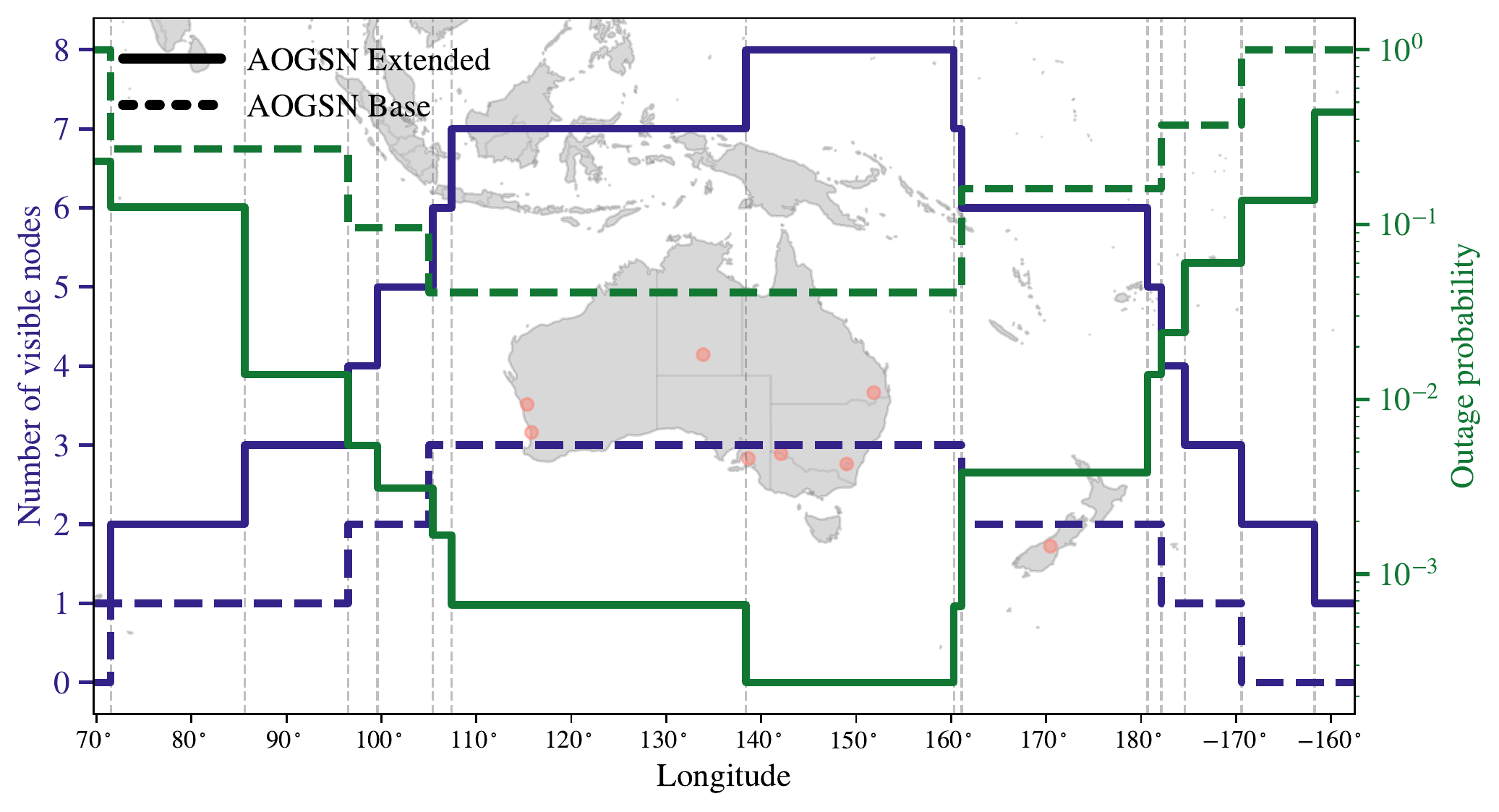}
  \caption{Number of visible AOGSN (extended, solid line; base, dashed line) nodes (blue) and corresponding outage probability (green), $p_\text{avail}(M=0)$, for GEO satellites as a function of longitude. GEO visibility to AOGSN ground stations (orange) is with $\theta_\text{elevation}>30^\circ$. Outage probabilities are estimated by sampling the joint probability distribution of site availability, as in \autoref{sec:div}.}
  \label{fig:geovisibility}
\end{figure*}

\section{GEO visibility analysis}\label{sec:geo_visibility}

While we have extensively studied the case of how the AOGSN can support LEO satellites, a large and diverse optical network can provide very high reliability to GEO or deep space if there is significant mutual visibility to numerous OGS. GEO optical feeder links to ground may become an important future application of FSOC networks, where ground segment reliability is critical~\cite{bonnefois2019adaptive,vedrenne2019first}. The Asia-Pacific region may soon host a GEO satellite with an FSOC terminal to demonstrate this capability, which an Australasian network could reliably support~\cite{kolev2022status,carrasco2022development}.

We therefore use orbital simulations with \textit{a.i.-solutions}' FreeFlyer to determine GEO visibility to each ground station in the AOGSN with $\theta_{elevation}>30^\circ$. For simplicity, GEO orbital placement is considered only a function of longitude with $h_\text{orbit}=35,786$~km and $i\lesssim1^\circ$. Hence, GEO longitude corresponds to visibility with $N\leq 8$ nodes, allowing estimates of $p_\text{avail}(M=0)$, described in \autoref{sec:div}, as a function of longitude. \autoref{fig:geovisibility} shows the number of nodes in the extended AOGSN visible to a GEO as a function of longitude. A portion of the AOGSN is therefore visible to a GEO between different longitude bounds. For each segment, e.g. $N=8$ from $140^\circ$  to $160^\circ$, \autoref{fig:geovisibility} shows $p_\text{avail}(M=0)$ for that set of sites. The extended AOGSN is three times longer in longitude than latitude, which helps maintain $p_\text{avail}(M=0)<0.01$ over $\approx80^\circ$ of longitude as shown in \autoref{fig:geovisibility}. This example network could then feasibly provide reliable support to GEO satellites over nearly $1/4$ of the planet. The low number of visible nodes to the East and West of Australia in \autoref{fig:geovisibility} indicates extensions in these directions are ideal if minimising GEO communication outages are a high priority. Placing an OGS in the Polynesian region, e.g. Tahiti, could be an ideal extension by allowing visibility for North American GEO satellites such as the Laser Communications Relay Demonstration~\cite{israel2017laser}.
\autoref{fig:geovisibility} further highlights the improvements between the base and extended AOGSN.

\section{Conclusion}\label{sec:conclusion}

\subsection{Summary and key results}
We have developed and shown use cases for a number of novel methods for assessing and optimising FSOC ground segment networks, focusing, but not limited to, the availability and site-wise correlations within the Australasian region. An example network called the AOGSN is proposed as a realistic representation of an Australasian network, with base ($N=3$) and extended ($N=8$) variants. We furthermore perform detailed case studies on the base and extended AOGSN, highly-relevant to the future of optical communications within the Australasian region, comparing it to theoretically optimised networks throughout the study. 

For determining availability statistics of the networks, we compare between a range of remote sensing data from five Earth observation satellite sources, but particularly focus on the highly spatially and temporally resolved cloud cover data from the Himawari-8 GEO satellite. We model the cloud cover as a Bernoulli process, which provides us a mathematical formalism for the site availability, and joint availability probability distribution function, which we construct and sample to access outage probabilities. We provide a detailed analysis of the spatially resolved correlations between pair-wise ground stations, and between ground stations and arbitrary sites. These statistics are combined with orbital simulations to estimate link duration and network capacity statistics. We list the key results below:
\begin{itemize}
    \item We estimate a $69\%$ average site availability, and an average correlation of $\rabs=0.088$, for the extended AOGSN (comparable correlations to an ideal, optimised network; see \autoref{tab:networkcomp}), demonstrating the large (spatially uncorrelated) and arid nature of the Australasian region.
    \item Using site availability and correlation measurements and the method outlined in \cite{Macke2009_neuronal_spike_model}, in \autoref{sec:div} we construct the network availability probability distribution function for an arbitrary sized Australasian network, with arbitrary correlations between sites. The analysis of this correlated network shows the extended AOGSN has $p_\text{avail}(M=0)=2.4\times10^{-4}$, i.e. at any given time there is close to a one-in-five thousand chance of suffering a network-wide outage (shown in \autoref{fig:outageprobcurves}, alongside statistics for the AOGSN base and ideal, optimised networks), hence a large and diversified network such as the AOGSN is reliable enough to overcome the reliability challenge for FSOC space-to-ground communication. 
    \item We develop a spatially-resolved method for optimising the positions of ground stations based on the network availability and diversity, without site preselection. This method can be used to develop new or extend existing networks, which we show in \autoref{sec:optimisation} and \autoref{sec:roimasking}, respectively, and to generate an optimal network to provide an ideal reference for commissioned networks (e.g., for comparing with the preselected AOGSN). 
    \item In \autoref{sec:coverage} we combine orbital LEO simulations with previous our availability models to estimate network capacity as a function of orbital inclination, showing that the extended AOGSN network is close to optimal (compared to our ideal, optimised network) and both the AOGSN and optimal networks are capable of supporting bitrates of $\gtrsim 10$~Tb/day to a LEO.
\end{itemize}

\subsection{Limitations of our study and future work}

In this study we consider only a simple availability model based on time-averaged cloud cover. Clearly, as shown in \autoref{fig:aogsn_seasonal}, there is seasonal variation (which can lead up to a factor of $\approx 2$ variation in the availability) and temporal correlation that we do not account for in our study. We will leave a detailed analysis of how the performance of FSOC networks in the Australasian network will change with seasonal, climatic and temporal correlations for future work. Cirrus clouds, and other thin icy clouds, could be more rigorously discarded from the total cloud fraction, as they usually do not create an outage. A method for differentiating the liquid cloud fraction, i.e. approximately thick clouds, from the total cloud fraction is outlined in \cite{listowski2019xmltex} and could be adapted in future studies that want to add additionally complexity to the models we have presented.

Furthermore, the atmospheric channel will also constrain availability, which has been explored theoretically in other network diversity analyses~\cite{lyras2018optimum,erdogan2021site}. Very arid and flat sites with low cloud cover likely have strong and detrimental turbulence conditions, hence, the results presented in our study represent an upper bound of the site availabilities. This consideration may be especially important in the context of sensitive coherent protocols, which may require adaptive optics \cite{gregory2013three}. Network degradation due to turbulence will be highly system specific, so estimating system performance should be done on a case-by-case basis. We plan to deploy atmospheric turbulence instruments in prospective sites as part of future investigations to explore OGS placement for the Australasian region, with a focus on the suitability of low cloud cover desert and arid sites. However, for a network model that retains spatial freedom and without an experimental campaign at selected sites, there is no feasible or reasonable way to include turbulence in the present, spatially resolved outage model. Well-validated spatial estimates of daytime turbulence do not exist and testing emerging models (e.g. \cite{basu2020mesoscale,osborn2018atmospheric}), will require extensive experimental data for future models that include these effects. 

Our optimisation method described in \autoref{sec:optimisation} relies upon only two site characteristics: (1) the availability and (2) the spatially-resolved cloud cover correlation, both of which are equally weighted and without any spatial (latitude, longitude or altitude) constraints. The latitude-correction case studied in \autoref{sec:latitudecorr} of \autoref{sec:coverage} demonstrates the flexibility of this method by optimising for a specific orbit. The theoretical outage model involving the atmospheric channel developed in \cite{erdogan2021site} highlights that our cloud-only model, and proceeding literature, can only provide an upper bound of availability by not including turbulence-induced outages \cite{fuchs2015ground}.

\section*{Funding}
M.~B.~acknowledges funding from the Australian Government Research Training Program and the CSIRO postgraduate support scholarship. J.~R.~B.~acknowledges financial support from the Australian National University, via the Deakin Ph.D and Dean's Higher Degree Research (theoretical physics) Scholarships, the Australian Government via the Australian Government Research Training Program and the Australian Capital Territory Government funded Fulbright scholarship. N.~J.~R.and J.~E.~C~are supported by a New Zealand-DLR Joint Research Programme funded by New Zealand's Ministry of Business, Innovation and Employment.

\section*{Acknowledgements}
The \textit{Terra}/MODIS, \textit{Aqua}/MODIS, Suomi NPP/VIIRS, and \textit{Aqua}/AIRS datasets were acquired from the Level-1 and Atmosphere Archive \& Distribution System (LAADS) Distributed Active Archive Center (DAAC), located in the Goddard Space Flight Center in Greenbelt, Maryland. Himawari-8 data used for this paper has been provided by the Earth Observation Research Centre of the Japan Aerospace Exploration Agency. Data from the Modern-Era Retrospective analysis for Research and Applications, Version 2 (MERRA-2) was provided by the National Aeronautics and Space Administration Global Modelling and Assimilation Office, located in the Goddard Space Flight Center in Greenbelt, Maryland. We also thank the anonymous reviewers who helped enhance the clarity and presentation of our study.\\

We acknowledge the open-source data analysis and visualisation software used in this study: \texttt{numpy} \citep{Oliphant2006,numpy2020}, \texttt{matplotlib} \citep{Hunter2007}, \texttt{scipy} \citep{virtanen2020scipy}, \texttt{cartopy} \citep{Cartopy}.

\bibliography{main}

\appendix
\section*{Appendix}

\section{Optimisation for additional sites and regions}
\label{sec:roimasking}
\begin{figure}[!ht]
  \centering
  \includegraphics[width=\linewidth]{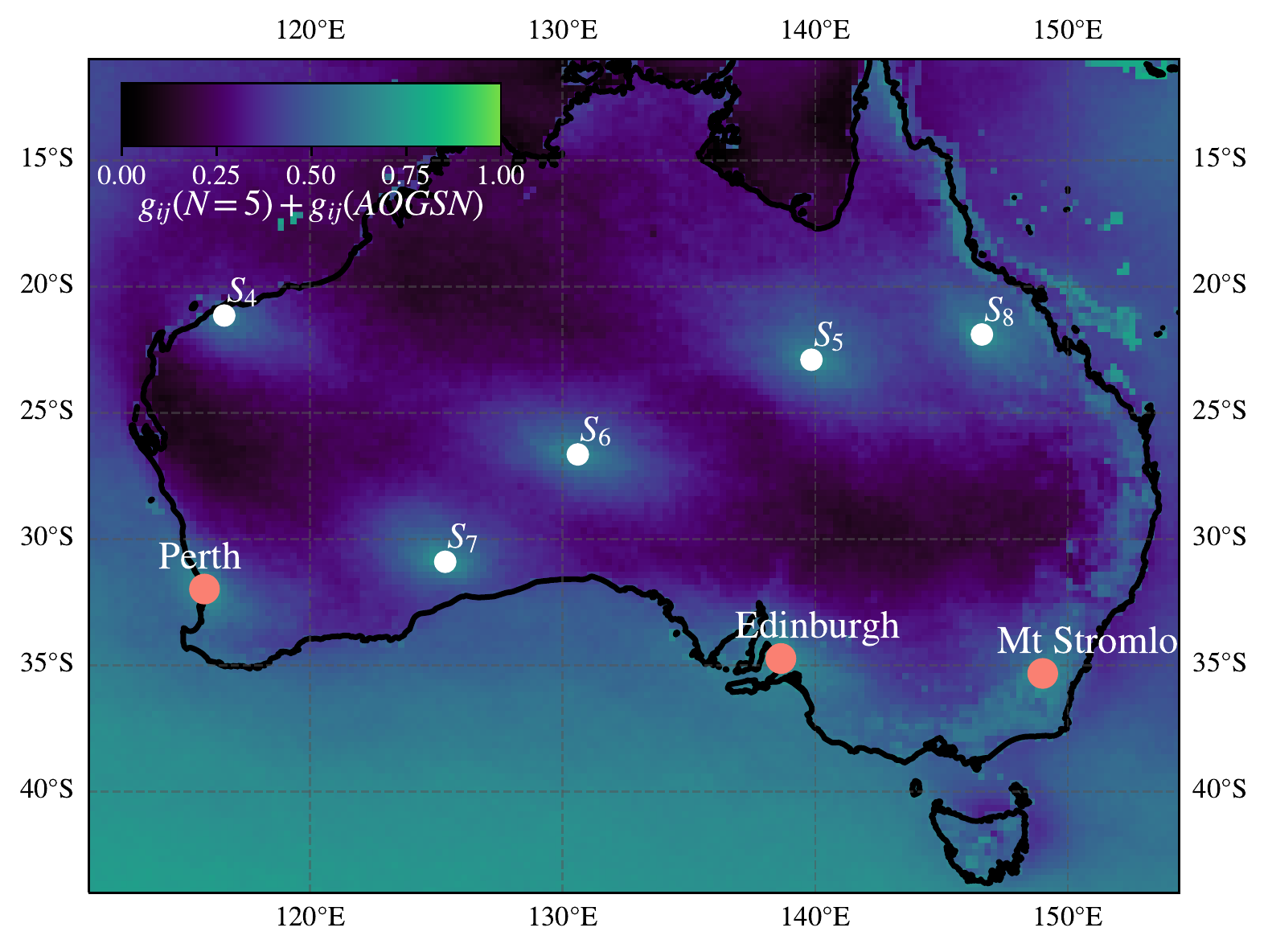}
  \caption{The same as \autoref{fig:beattiemap} but for $N=5$ with the base AOGSN ground stations as initial conditions. Daily AHI/Himawari-8 images from 2015-2022 are used for determining the combined $g_{ij}(N=5)+g_\text{AOGSN}(N=3)$ optimisation function.}
  \label{fig:aogsnstartalgo}
\end{figure}

Our optimisation method described in \autoref{sec:optimisation} can be used to select additional sites for an existing network. This is done by first constructing the $g_{ij}(N)$ function for the initial ground stations and then adding sites following the process used to construct $\N_{N=8}$ and $\mathcal{M}_{N=8}$. We test this using the AOGSN base network as the initial condition, and run the algorithm until it reaches $N=8$, as shown in \autoref{fig:aogsnstartalgo}. This method could be particularly useful when combined with the capacity estimation from \autoref{sec:coverage}, as the improvement from optimally adding a new OGS to an existing network can be quantified. This application could also be useful with additional ROI masking functions to guide site selection for extending a given network, such as the AOGSN. \autoref{fig:aogsnstartalgo} indicates that for optimisation, the best site to extend the base AOGSN with would be located in northern Western Australia.

\begin{figure}[!ht]
  \centering
  \includegraphics[width=\linewidth]{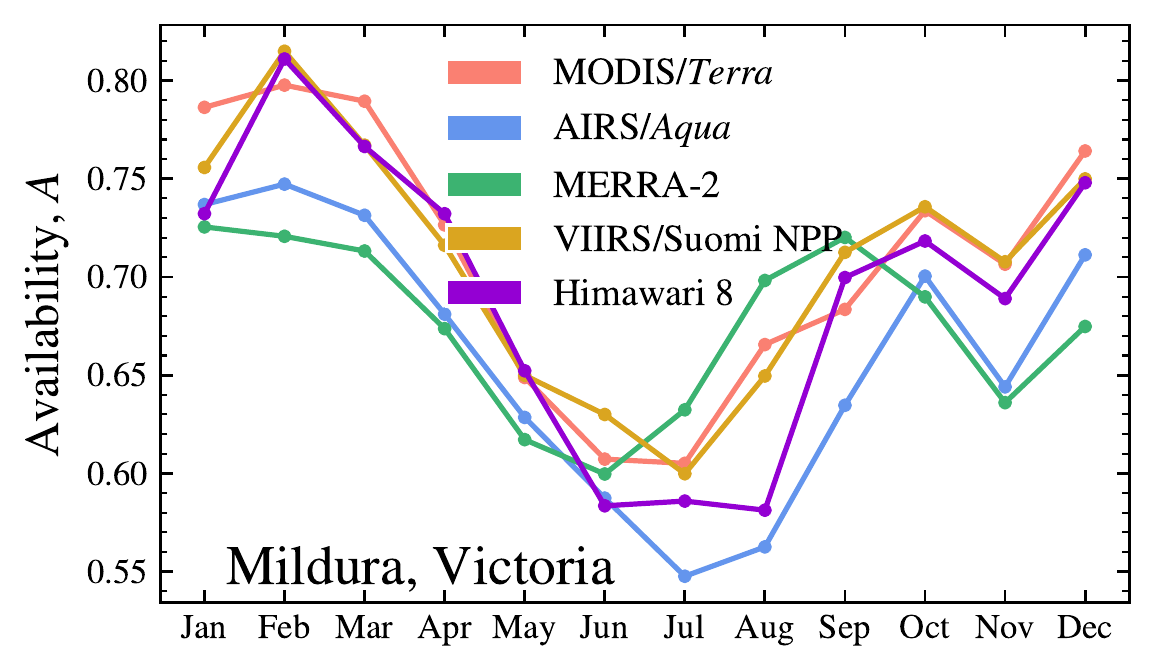}
  \caption{Availability as a function of month for Mildura AOGSN site, showing the seasonal variability and each separate cloud retrieval product (different colours) used in this paper, which are averaged over in \autoref{fig:aogsn_seasonal}.}
  \label{fig:mildura_seasonal}
\end{figure}

\section{Remote sensing satellite variability}\label{sec:satellitevariation}
In \autoref{sec:cover} we outlined the method of using all five satellites from \autoref{tab:sats} to estimate $A$ (or $\Omega$) in a robust fashion. \autoref{fig:mildura_seasonal} shows the different seasonal $A$ curves from all five data products. We can see clearly from \autoref{fig:mildura_seasonal} that there is reasonable agreement between all satellites despite different methods and resolution. However, Alice Springs suffered issues with Himawari and MERRA-2, while Mt John had issues with \textit{Aqua}/AIRS and MERRA-2 due to snow and ice reflectance. In these cases, those data streams were discarded as outliers. From \autoref{fig:mildura_seasonal}, we can see that higher resolution data, Himawari-8, finds agreement with low resolution data, indicating that spatial averaging is often appropriate. The reliability of spatial averaging owes to Australia's relatively homogeneous terrain, and could cause errors in other environments. High resolution swathes from the \textit{Aqua}, \textit{Terra}, and Suomi NPP LEO satellites can be used if the extra resolution is required. 

\begin{figure}[!ht]
  \centering
  \includegraphics[width=\linewidth]{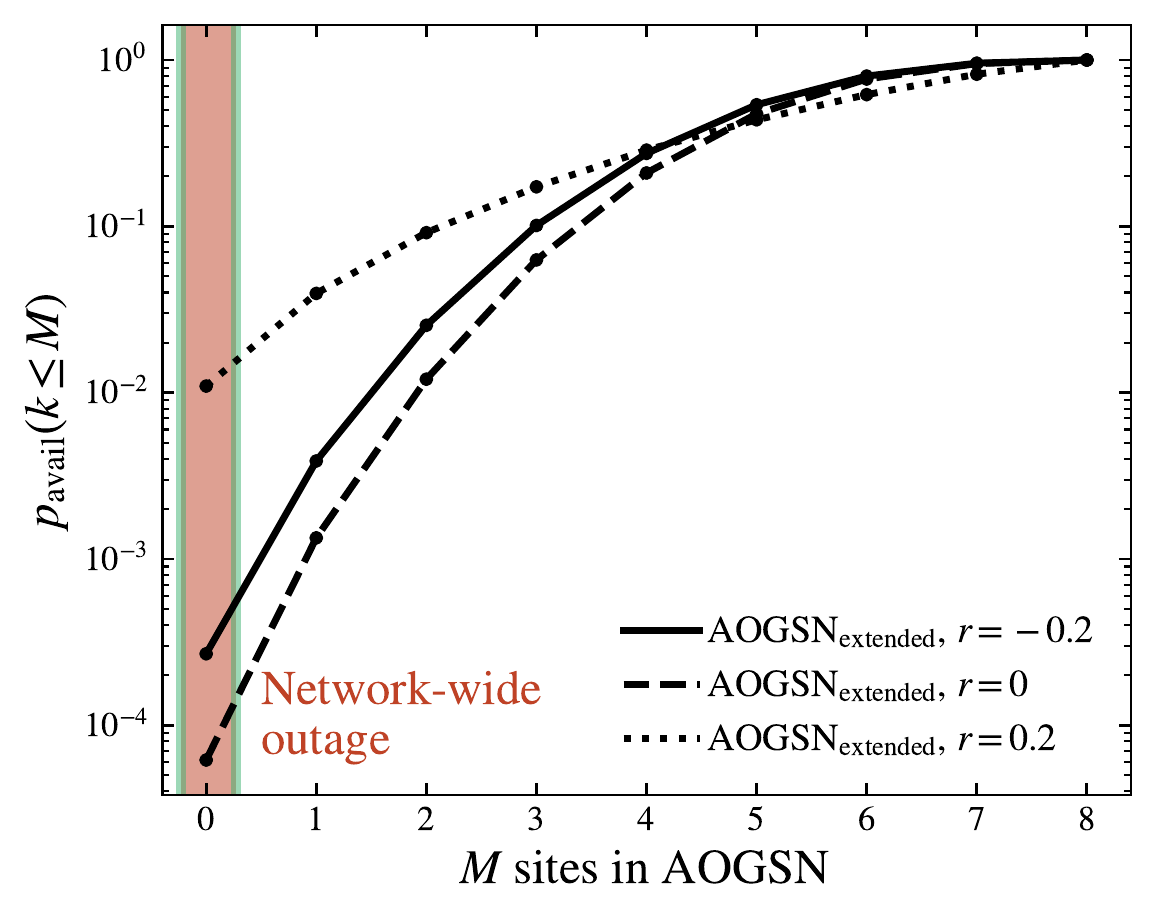}
  \caption{The same as \autoref{fig:outageprobcurves} but for three networks which admit to the AOGSN extended configuration with negative correlation (solid; $r=-0.2$), no correlation (dashed; $r=0.0$) and positive correlation (dotted; $r=0.2$) between ground stations.}
  \label{fig:anticorrelations}
\end{figure}

\section{A network with negative correlations between ground stations}
\label{sec:anticorroutages}

The most prominent negative correlations we observe in Figure \ref{fig:corr_matrix} are due to seasonal inversions. Negative correlations between network nodes intuitively boosts diversity, i.e. in the $N=2$ case the minima of their availability curves are out of phase. We test how negative correlations influence the network statistics by constructing cumulative distributions functions, as in \autoref{sec:div} for a negatively correlated network $(r = -0.2)$, an uncorrelated network $(r=0.0)$, and positively correlated network $(r=0.2)$, using the AOGSN extended ground station configuration. We show the results in \autoref{fig:anticorrelations}, where the $r = -0.2$ is shown with the solid line, $r = 0.0$ with the dashed, and $r = 0.2$ with the dotted. We find that all networks have the same behaviour for large $M$, as we showed previously in \autoref{fig:outageprobcurves}, and discussed in \autoref{sec:div}. However, for $M=0$ to $M\leq 4$, the networks show different availability statistics. For this range of $M$, the network without correlations performs the best, followed by the negatively correlated network and then the positively correlated network. This shows that negatively correlated networks admit to better availability statistics than positively correlated networks, but networks without correlations between ground stations perform the best.

\end{document}